\documentclass[a4paper,12pt,english]{article}
\usepackage[T1]{fontenc}
\usepackage[latin1]{inputenc}
\usepackage{cite}
\usepackage{epsfig}
\usepackage{amsmath,graphics,bbm,amssymb}
\usepackage{babel}

\newcommand{\op}[1]{#1}
\begin{document}

\title{\begin{flushright}
{\normalsize IPPP/03/60\\
DCPT/03/120\\}
\end{flushright}\vspace{1cm}
$N$-point Amplitudes in Intersecting Brane Models}
  \author{S.A.Abel and A.W.Owen}
 \date{\textit{IPPP, Centre for Particle Theory, Durham
  University, DH1 3LE, Durham, U.K.}}
  
\maketitle

\begin{abstract}
\noindent We derive general and complete expressions for 
$N$-point tree-level amplitudes in Type II string models
with matter fields localised at D-brane intersections.

\end{abstract}

\section{Introduction}

Since their discovery, and their subsequent relation to gauge theories,
D-branes have become established as a central element in the development of
phenomenologically viable string models. They facilitate immediate and
straightforward construction of models with interesting gauge groups. 
However, the requirement of chirality in any physically realistic model leads
to a restricted number of possible D-brane set-ups. An important class are
the intersecting brane models~\cite{ralph1,ibws}. This scenario
exploits the fact that chiral fermions can arise at the intersection of two
branes at angles~\cite{bangles}, hence the spectrum of fermions is
determined by the intersection numbers of the D-branes which wrap some compact
space. As a consequence, a simple and rather attractive topological
explanation of family replication is obtained. 

The intersecting brane scenario has been remarkably successful in producing
semi-realistic models, for example,
Refs.~\cite{ralph1,chiral,ralph2,ralph4,ralph5,ralph6,intflip,smcomp,moresm,bai,hugetitle,ralph8,quasi,flip}.
Models similar to the Standard Model can be
obtained~\cite{just,ralph3,just2,ckok1,exsm2,cvetic3,exsm1} and furthermore viable
constructions with $N=1$ supersymmetry have been developed~\cite{cvetic1,cvetic2,susy,susyorient,cvetic4,cvetic5,cvetic6,localsusy}, although this latter possibility is more difficult to achieve.
A more detailed analysis of the phenomenology of such models is currently
in progress, for example~\cite{deform,gthresholds,witt,ralph7}. In particular, computations
of Yukawa couplings~\cite{qmass,yukawa} and flavour changing neutral
currents~\cite{ams} have been performed.

In this paper we generalise the methods used for the computation of
the three and four point amplitudes in~\cite{mirjam,paper2} to the case of
$N$-point amplitudes, In particular, deriving complete expressions for both the
classical and quantum contributions.  

Our analysis will be based on the technology discussed in~\cite{hamidi,dixon,atick} for closed strings on orbifolds. This is due to an
analogy between twisted closed string states on orbifolds and open strings at
brane intersections. This analogy has been discussed previously
in ref.~\cite{paper2} and will be briefly recapped in our first section. We
then proceed to a determination of the classical part of the general $4$-point
amplitude. This calculation is then generalised to the N-point case along
with a prescription for obtaining the wrapped contributions to the
amplitude in the case of toroidal geometry. Next, we use conformal field theory techniques to
evaluate the quantum contribution to a general four point amplitude with
three independent angles. This calculation is then fully generalised in the
determination of the
quantum contribution to the $N$-point tree level amplitude. As a consistency
check, we proceed to demonstrate that the $(N-1)$-point amplitude can be
obtained as a limiting case of the $N$-point amplitude.

Finally, as an example of the application of our results, we analyse the situation of four independent sets of
branes. In particular, we discuss the relevant process $q_{L}q_{R}
\rightarrow e_{L}e_{R}$ which, depending on the D-brane geometry, is a purely
stringy effect, or has a field theory limit corresponding to a Higgs
exchange. In the latter case, we extract the Higgs pole and determine the mass of the Higgs
purely from the four point amplitude. Throughout this work we will focus on the case of D6-branes wrapping the compact internal space
$T^{2} \times T^{2}\times T^{2}$, however, our results are easily adapted to
other cases.
  
\section{Closed and open string twisted states}

Let us begin by briefly reviewing the analogy between closed strings on
oribfolds and open strings at D-brane intersections. This analogy allows us
to employ the CFT techniques developed in~\cite{dixon,hamidi}, which forms
a basis for our analysis. 

\begin{figure}
\begin{center}
 \epsfig{file=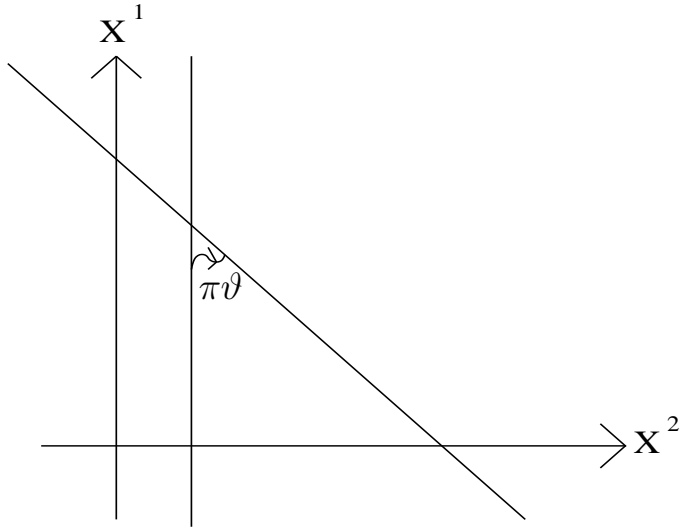,height=70mm,width=90mm}
\put(-183,87){\large{$\pi\vartheta$}}
\caption{A `twisted' open string state}
\label{stringpic}
\end{center}
\end{figure}

An open string stretched between two D-branes intersecting at an angle $\pi
\vartheta$, as depicted in figure~\ref{stringpic}, has the boundary conditions,
\begin{equation}
\label{bc}
\begin{array}{l}
\partial_{\tau}X^{2}(0)=\partial_{\sigma}X^{1}(0)=0, \\
\partial_{\tau}X^{1}(\pi)+\partial_{\tau}X^{2}(\pi)\cot(\pi\vartheta)=0, \\
\partial_{\sigma}X^{2}(\pi)-\partial_{\sigma}X^{1}(\pi)\cot(\pi\vartheta)=0.
\end{array}
\end{equation}
Thus we can determine the holomorphic solutions to the string equation of motion to be,
\begin{equation}
\label{modeexp}
\begin{array}{l}
\partial X(z)=\sum_{k}\alpha_{k-\vartheta}z^{-k+\vartheta-1},\\
\partial \bar{X}(z)=\sum_{k}\bar{\alpha}_{k+\vartheta}z^{-k-\vartheta-1},
\end{array}
\end{equation}
where $z=-e^{\tau-i\sigma}$ is the worldsheet coordinate and has domain the
upper-half complex plane. This domain is extended to the entire complex
plane using the `doubling trick', i.e. we define,
\begin{equation}
\partial X(z)= \left\{ \begin{array}{ll}
                         \partial X(z) & \mbox{Im}(z) \geq 0 \\
                         \bar{\partial}\bar{X}(\bar{z}) & \mbox{Im}(z) <
                         0
                        \end{array} \right.,
\end{equation}
and similarly for $\partial \bar{X}(z)$.

Now the mode expansion of a closed string state in the presence of a
$\mathbb{Z}_{N}$ orbifold twist field, is identical to~(\ref{modeexp}) with the
replacement $\vartheta=\frac{1}{N}$. Hence, we see that there is a natural
correspondence between open strings stretched between intersecting branes and a
twisted closed string state on an orbifold. To take account of this
correspondence, we must introduce a twist field $\sigma_\vartheta
(w,\bar{w})$ for the open string, this field changes the boundary conditions of
$X$ to be those of eq.(\ref{bc}), where the intersection point of the two
D-branes is at $X(w,\bar{w})$. Proceeding as in the closed string
case we obtain the OPEs,
\begin{equation}
\label{opes}
\begin{array}{l}
\partial X(z) \sigma_{\vartheta}(w,\bar{w}) \sim
(z-w)^{-(1-\vartheta)}\tau_{\vartheta}(w,\bar{w}), \\
\partial \bar{X}(z) \sigma_{\vartheta}(w,\bar{w}) \sim
(z-w)^{-\vartheta}\tau'_{\vartheta}(w,\bar{w}),
\end{array}
\end{equation}
where $\tau'_{\vartheta}$ and $\tau_{\vartheta}$ are excited twist
fields. The local monodromy conditions for transportation around
$\sigma_{\vartheta}(w,\bar{w})$ are simply determined from the OPEs to be,
\begin{equation}
\label{mono}
\begin{array}{l}
\partial X(e^{2 \pi i}(z-w))=e^{2\pi i\vartheta}\partial X(z-w), \\
\partial \bar{X}(e^{2 \pi i}(z-w))=e^{-2\pi i\vartheta }\partial \bar{X}(z-w).
\end{array}
\end{equation}
The mode expansion for $X$ is then,
\begin{equation}
X(z,\bar{z})=\sqrt{\frac{\alpha'}{2}}\sum_{k}\left( \frac{\alpha_{k-\vartheta}}{k-\vartheta}z^{-k+\vartheta}+\frac{\tilde{\alpha}_{k+\vartheta}}{k+\vartheta}\bar{z}^{-k-\vartheta}\right),
\end{equation}
with the right and left moving modes being mapped into upper and lower
half planes. A similar mode expansion is obtained for the fermions with
the obvious addition of \( \frac{1}{2} \) to the boundary conditions
for NS sectors.  

\begin{figure}
\begin{center}
 \epsfig{file=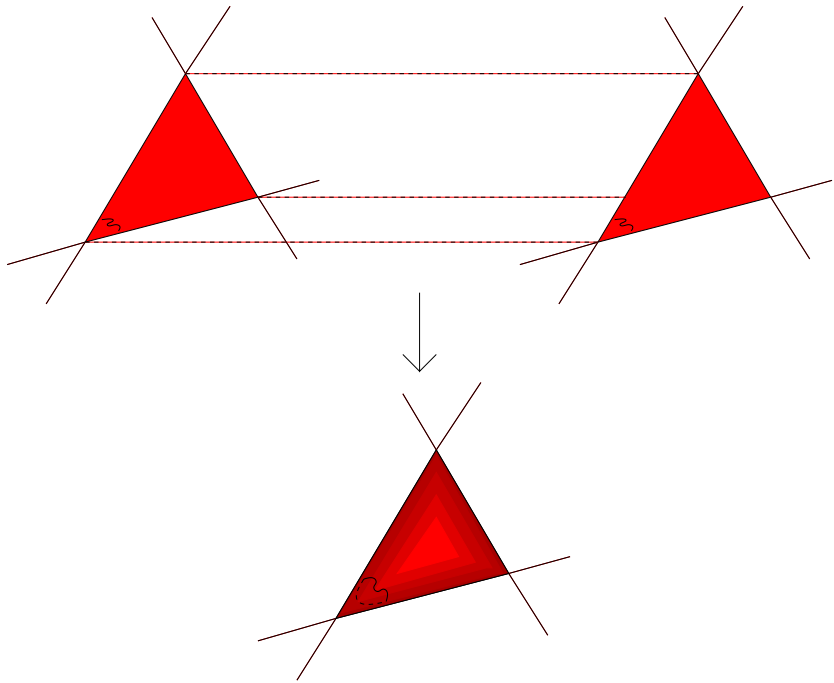,height=90mm,width=110mm}
\caption{Identifying open strings to form closed strings}
\label{sqrtpic}
\end{center}
\end{figure}

The correspondence with twisted states on orbifolds (or conifolds) can be
seen geometrically as in figure~\ref{sqrtpic}. This figure shows two
identical three point diagrams which are sewn together at their edges. An open
string living at the intersection is doubled up to form a twisted closed
string. As a result we expect to recover the famous $\mbox{open string}
\sim \sqrt{\mbox{closed string}}$ relation. However, we also note that the
intersection angles in this case are more general than the rather restrictive
ones found in supersymmetric orbifolds of closed strings.

\section{The classical part of the $N$-point function}
We now proceed to an analysis of the $N$-point function for
string states localised at $N$ distinct D-brane intersections. We assume the
$N$ D-branes form the boundary of an $N$-sided polygon, with vertices $f_{i}$
and interior angles $\pi \vartheta_{i}$ such that,
\begin{equation}
\sum_{i=1}^{N}\vartheta_{i}= N-2. 
\end{equation}
The D-brane intersecting $f_{i}$ and $f_{i+1}$ is labelled the
$i^{th}$ D-brane. We will consider two distinct cases for our compactification
space M, 
\begin{itemize}
\item planar, $M=\mathbb{R}^2\times\mathbb{R}^2\times\mathbb{R}^2 $,
\item and toroidal, $M=T_{2} \times T_{2} \times T_{2}$.
\end{itemize}
The planar case is useful for illustrating the methods used in the
computation without adding extra complication, and furthermore, will turn out to give the
leading order contribution in the more realistic toroidal scenario. Due to
our choice of $M$, 
the amplitude can be factorised into a product of three
identical contributions, one from each of the three subfactors
(i.e. $\mathbb{R}^2$ or $T_{2}$). In the following, we will therefore
concentrate on the contribution from one such subfactor.

We denote the spacetime coordinates for a particular subfactor by
$X=X^{1}+iX^{2}$ and $ \bar{X}=X^{1}-iX^{2}$. The bosonic
field $X$ can be split up into a classical piece, $X_{cl}$, and a quantum
fluctuation, $X_{qu}$. The amplitude then
factorises into classical and quantum components,
\begin{equation}
Z=\sum_{\langle X_{cl} \rangle}e^{-S_{cl}}Z_{qu},
\end{equation}
where,
\begin{equation}
\label{stringaction}
S_{cl}=\frac{1}{4\pi \alpha}\left( \int d^{2}z(\partial X_{cl}
  \bar{\partial}\bar{X}_{cl}+\bar{\partial} X_{cl} \partial \bar{X}_{cl}) \right).
\end{equation}
$X_{cl}$ must satisfy the string equation of motion and possess the correct
asymptotic behaviour near the polygon vertices. 

\subsection{Determining the classical $4$-point function for the plane}
\label{solnsandmono}
Before discussing the calculation of the classical contribution to the
general $N$-point amplitude, let us first consider the simpler four point case.
The $4$-point function requires $4$ twist operators, 
$\sigma_{\vartheta_{i}}(z_{i},\bar{z}_{i})$, one for each polygon
vertex or D-brane intersection. These twist operators are attached to the boundary of
the tree-level disc diagram which can be conformally mapped to the upper-half
plane. Hence, using $SL(2,\mathbb{R})$ invariance we set $z_{1}=0$, $z_{2}=x_{2}$,
$z_{3}=1$ and $z_{4}=x_{\infty}$. The classical solution, $\partial X_{cl}$, is determined up to a
normalisation constant by its holomorphicity and asymptotic 
behaviour at each D-brane intersection,
which is given by the OPEs~(\ref{opes}). Thus,
\begin{equation}
\label{classsolns}
\begin{array}{ll}
\partial X_{cl}(z)= a\omega(z), & \partial \bar{X}_{cl}(z)=\bar{a}
\omega'(z), \\
\bar{\partial} X_{cl}(\bar{z})= b\bar{\omega}'(\bar{z}), & \bar{\partial} \bar{X}_{cl}(\bar{z})=\bar{b}\bar{\omega}(\bar{z}),
\end{array}
\end{equation}
where,
\begin{equation}
\begin{array}{ll}
\omega(z)=\prod_{i=1}^{4} (z-x_{i})^{-(1-\vartheta_{i})} &
\omega'(z)=\prod_{i=1}^{4} (z-x_{i})^{-\vartheta_{i}},
\end{array}
\end{equation}
and $a,\bar{a},b,\bar{b} \in \mathbb{C}$ are normalisation constants. Using the
doubling trick to define $\partial X_{cl}$ on the entire complex plane,
 we require $a^{*}=\bar{b}$ and $\bar{a}=b^{*}$ (upto a phase).

The normalisation of the classical solutions are determined from the global
monodromy conditions, i.e. the transformation of $X$ as it is transported
around more than one twist operator, such that the net twist is zero. We determine this from the action of a
single twist operator, $\sigma_{\vartheta}(w,\bar{w})$, which in the planar
case acts to transform $X(z,\bar{z})$ as,
\begin{equation}
\label{trans}
X(e^{2\pi i}z,e^{-2\pi i}\overline{z})= e^{2\pi i\vartheta}X +(1-e^{2\pi i\vartheta})f,
\end{equation}
where f is the intersection point of the two D-branes. This can be seen from
the local monodromy conditions~(\ref{mono}) and the fact that $f$ must be left
invariant. The global monodromy of $X$ is then simply a product of such
actions. Note, if we split $X$ up into a classical and quantum part, then the
classical field should have exactly the same behaviour as the full field.
Hence, the boundary conditions for $X_{qu}$ ignores the shift $(1-e^{2\pi
  i\vartheta})f$ leaving,
\begin{equation}
\label{transqu}
X_{qu}(e^{2\pi i}z,e^{-2\pi i}\overline{z})= e^{2\pi i\vartheta}X_{qu}.
\end{equation}

\begin{figure}
\centering
\epsfig{file=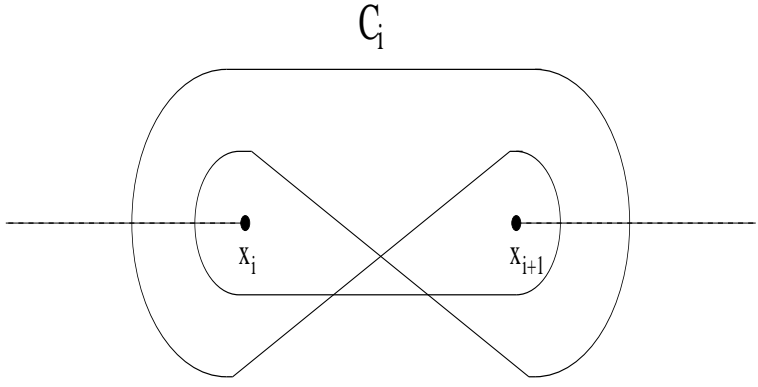, height=50mm, width=100mm}
\caption{The Pochammer loop $C_{i}$ with net twist zero, branch cuts lie
  along the dashed lines.}
\label{curve}
\end{figure}

For the $4$-point case, there exists $2$ independent contours, $C_{i}$, $i=1,2$, with net twist
zero. Each $C_{i}$ is a Pochammer loop, which is illustrated in
figure~\ref{curve}. The $2$ independent global monodromy conditions for the
classical field are,
\begin{equation}
\label{monocond}
\Delta _{C_{i}}X_{cl}=e^{-\pi i(\vartheta_{i}-\vartheta_{i+1})}4\sin (\pi \vartheta_{i})\sin (\pi\vartheta_{i+1})(f_{i+1}-f_{i})=\oint _{C_{i}}dz\partial X_{cl}(z)+\oint_{C_{i}}d\overline{z}\bar{\partial}X_{cl}(\bar{z}).
\end{equation}
The contour integrals required can be written as,
\begin{equation}
\label{integrals}
\begin{array}{l}
\oint_{C_{i}} dz \omega(z)=e^{-\pi i(\vartheta_{i}-\vartheta_{i+1})}4\sin (\pi \vartheta_{i})\sin (\pi
\vartheta_{i+1})F_{i}, \\
\oint_{C_{i}} d\bar{z} \bar{\omega}'(\bar{z})=e^{-\pi i(\vartheta_{i}-\vartheta_{i+1})}4\sin (\pi \vartheta_{i})\sin (\pi
\vartheta_{i+1})\bar{F}'_{i},
\end{array}
\end{equation}
where,
\begin{equation}
\begin{array}{ll}
F_{i}=\int_{x_{i}}^{x_{i+1}}\prod_{j=1}^{4} (y-x_{j})^{-(1-\vartheta_{j})}dy
& F'_{i}=\int_{x_{i}}^{x_{i+1}}\prod_{j=1}^{4} (y-x_{j})^{-\vartheta_{j}}dy.
\end{array}
\end{equation}
Hence~(\ref{monocond}) can be simplified to,
\begin{equation}
\label{monosimple}
\begin{array}{ll}
aF_{i}+b\bar{F}'_{i}=f_{i+1}-f_{i} & i=1,2.
\end{array}
\end{equation}
Using these conditions we can determine a and b. Construct the classical solutions
$\partial X_{cl,1}$ and $\partial X_{cl,2}$, with
coefficients $\alpha_{i},\beta_{i}$ as in~(\ref{classsolns}) and which have the
simple global monodromy,
\begin{equation}
\label{simplemono}
\begin{array}{ll}
 \alpha_{i}F_{j}+\beta_{i}\bar{F}'_{j}=\delta_{ij}, & i,j=1,2.
\end{array}
\end{equation}
Then,
\begin{equation}
\label{cons}
\begin{array}{ll}
\alpha_{1}=\frac{\bar{F}'_{2}}{F_{1}\bar{F}'_{2}-F_{2}\bar{F}'_{1}}, &
\alpha_{2}=-\frac{\bar{F}'_{1}}{F_{1}\bar{F}'_{2}-F_{2}\bar{F}'_{1}}, \\
\beta_{1}=-\frac{F_{2}}{F_{1}\bar{F}'_{2}-F_{2}\bar{F}'_{1}}, & \beta_{2}=\frac{F_{1}}{F_{1}\bar{F}'_{2}-F_{2}\bar{F}'_{1}}.
\end{array}
\end{equation}
It follows, that to satisfy~(\ref{monosimple}) for $i=1,2$, we require,
\begin{equation}
\label{normalisation}
\begin{array}{l}
a=(f_{2}-f_{1})\alpha_{1}+(f_{3}-f_{2})\alpha_{2}, \\
b=(f_{2}-f_{1})\beta_{1}+(f_{3}-f_{2})\beta_{2}. 
\end{array}
\end{equation}

Finally, using~(\ref{stringaction}) and~(\ref{classsolns}) the classical
contribution to the $4$-point function in the planar case is given by,
\begin{equation}
\label{4ptcls}
S_{cl}(x_{2})=\frac{1}{4\pi \alpha'}\left(|a|^2 I(x_{2})+|b|^2 I'(x_{2})\right),
\end{equation}
where,
\begin{equation}
\begin{array}{ll}
I(x_{2})=\int d^{2}z|\omega(z)|^2 & I'(x_{2})=\int d^{2}z|\omega'(z)|^{2}.
\end{array}
\end{equation}
We now proceed to the generalisation of this calculation to the $N$-point case.

\subsection{Generalisation to the $N$-point function} 
\label{Nptclassical}
 The $N$-point function requires $N$ twist operators,
 $\sigma_{\vartheta_{i}}(x_{i})$, where $x_{1}=0$, $x_{N-1}=1$
 and $x_{N}=\infty$. The classical solution is again determined up to a
 normalisation constant by it's holomorphicity and asymptotic behaviour at
 each D-brane intersection. However, in this case we may generalise our
 expressions for the classical solutions to,
\begin{equation}
\label{gensolns}
\begin{array}{ll}
\partial X_{cl}(z)= a\omega(z), & \partial \bar{X}_{cl}(z)=\rho(z)\omega'(z), \\
\bar{\partial} X_{cl}(\bar{z})= \bar{\rho}(\bar{z})\bar{\omega}'(\bar{z}), & \bar{\partial} \bar{X}_{cl}(\bar{z})=a^{*}\bar{\omega}(\bar{z}),
\end{array}
\end{equation}
where $\rho(z)$ is a polynomial of degree $p$ and, 
\begin{equation}
\begin{array}{ll}
\omega(z)=\prod_{i=1}^{N} (z-x_{i})^{-(1-\vartheta_{i})} &
\omega'(z)=\prod_{i=1}^{N} (z-x_{i})^{-\vartheta_{i}}.
\end{array}
\end{equation}
This modification obviously does not change the asymptotics of $X_{cl}$ at the D-brane
intersections. However, if we consider the contribution to $S_{cl}$,
\begin{equation}
\int d^{2}z \bar{\partial}X_{cl}\partial \bar{X}_{cl}=\int d^{2}z
|\rho(z)|^{2}|\omega'(z)|^{2},
\end{equation} 
we see that this converges only if $p \leq N-4$. Notice that for $N=4$ the
expressions in~(\ref{gensolns}) reduce to those of our previous case
in~(\ref{classsolns}). We can define a useful basis for $\rho(z)$ given by,
\begin{alignat}{1}
\rho^{i}(z)=\prod_{\stackrel{\mbox{\scriptsize $j=2$}}{\mbox{\scriptsize $(j
      \neq i$)}}}^{N-2}(z-x_{j}),  & \hspace{0.5cm} i=2,..,N-2.
\end{alignat}
Our generalised classical solutions can then be expressed in the form,
\begin{equation}
\label{genanti}
\begin{array}{ll}
\bar{\partial}X_{cl}(\bar{z})=b_{i}^{*}\bar{\Omega}'^{i}(\bar{z}), & \partial \bar{X}_{cl}(z)=b_{i}\Omega'^{i}(z),
\end{array}
\end{equation}
where,
\begin{equation}
\label{newom}
\Omega'^{i}=\omega'(z)\rho^{i}(z).
\end{equation}

In the $N$-point case our classical solutions require $N-2$ normalisation
constants. However, we now have $N-2$ Pochammer loops,
$C_{l}$, $i=l,..,N-2$ to which we can apply the global monodromy
condition~(\ref{monocond}). The required contour integrals are given by the
$(N-2)\times (N-2)$ matrix $W^{i}_{l}$, defined by
\begin{equation}
\label{nints}
\begin{array}{l}
W^{1}_{l}=\oint_{C_{l}} dz \omega(z)=e^{- i
  \pi(\vartheta_{l}-\vartheta_{l+1})}4 \sin( \pi \vartheta_{l})\sin( \pi
  \vartheta_{l+1})F_{l}^{1}, \\
W^{i}_{l}=\oint_{C_{l}} d\bar{z} \bar{\Omega}'^{i}(\bar{z})=e^{- i
  \pi(\vartheta_{l}-\vartheta_{l+1})}4 \sin( \pi \vartheta_{l})\sin( \pi
  \vartheta_{l+1})F_{l}^{i},
\end{array}
\end{equation}
where,
\begin{equation}
\begin{array}{ll}
F_{l}^{1}=\int_{x_{l}}^{x_{l+1}}\omega(y)dy, & \\
F_{l}^{i}=\int_{x_{l}}^{x_{l+1}}\bar{\omega}'(y)\rho^{i}(y)dy, & i=2,..,N-2.
\end{array}
\end{equation}
Substituting into~(\ref{monocond}) we obtain the global monodromy conditions,
\begin{equation}
\label{genmono}
\begin{array}{ll}
c_{i}F^{i}_{l}=f_{l+1}-f_{l} & l=1,..,N-2,
\end{array}
\end{equation}
with $c_{i}=(a,b^{*}_{2},..,b^{*}_{N-2})$. So we now have $N-2$ conditions for $N-2$
constants. Thus,
\begin{equation}
c_{i}(x_{2},..,x_{N-2})=\sum_{l=1}^{N-2}(f_{l+1}-f_{l})(F^{-1})_{i}^{l}.
\end{equation}

Finally, we determine the classical contribution to the $N$-point function in
the planar case to be,
\begin{equation}
\label{classNpt}
S_{cl}(x_{2},..,x_{N-2})=\frac{1}{4\pi\alpha'}\left(|a|^{2}I+\sum_{i,j}b_{i}^{*}b_{j}I'_{\bar{i}j}
\right),
\end{equation}
where,
\begin{equation}
\begin{array}{ll}
I(x_{2},..,x_{N-2})=\int d^{2}z |\omega(x)|^{2}, & I'_{\bar{i}j}(x_{2},..,x_{N-2})=\int d^{2}z\bar{\Omega}'^{i}(\bar{z})\Omega'^{j}(z).
\end{array}
\end{equation}
Note that, for $N=4$ this reduces to~(\ref{4ptcls}).

\subsection{Determining the classical $N$-point function for the torus}
We next consider the more phenomenologically interesting case of
$M=T^{2}\times T^{2}\times T^{2}$ and show that our previous result is simply
the leading order contribution. Define, $T^{2}=\mathbb{R}^{2}/\Lambda$, where
$\Lambda$ is a lattice with basis vectors $v_{x}=R_{x}(1,0)$ and $v_{y}=R_{y}(0,1)$. We denote by $(n,m)$ a non-trivial
cycle winding $n$ times around the cycle defined by $v_{x}$ and $m$ times around
the cycle defined by $v_{y}$. The $i^{th}$ D-brane then wraps the 
cycle $(n_{i},m_{i})$ and the number of intersections with the $j^{th}$ D-brane
is given by the intersection number,
\begin{equation}
I_{ij}=n_{i}m_{j}-n_{j}m_{i}.
\end{equation}
These cycles gives rise to an infinite number of polygons which wrap the
torus and contribute to the N-point function. Our
computation will follow the same general methodology described above,
but with some important modifications. 

\begin{figure}
\centering
\epsfig{file=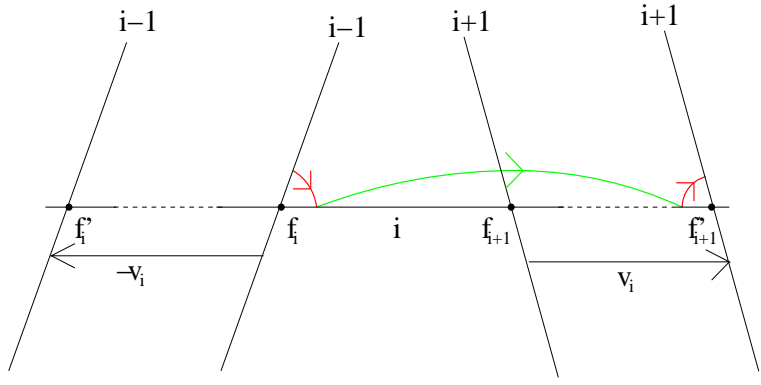, height=50mm, width=100mm}
\caption{Transporting $X_{cl}$ around two twist fields}
\label{shifts}
\end{figure}

On introduction of toroidal geometry, the action of our twist
operators must be generalised to not only rotate
$X_{cl}$ but also to shift by a lattice translation. Hence~(\ref{trans}) is
modified to
\begin{equation}
X(e^{2\pi i}z,e^{-2\pi i}\overline{z})= e^{2\pi i\vartheta}X +(1-e^{2\pi i\vartheta})(f+v),
\end{equation}
where $v \in \Lambda^{*}$, a coset of $\Lambda$. It follows, that on integrating around $C_{i}$, the portion of integration around each
vertex takes $X_{cl}(z,\bar{z})$ from one D-brane to another, while integrating
between the two vertices introduces a shift along the $i^{th}$ D-brane. Define $\vec{L}_{i,i+1}$ to lie in the direction
$f_{i+1}-f_{i}$, with magnitude the distance along the $i^{th}$ D-brane between
successive copies of the $(i+1)^{th}$ D-brane. Then the shift must be of the form $f_{i+1}-f_{i}+v_{i}$ where
$v_{i}=q_{i}|I_{i,i+1}|\vec{L}_{i,i+1} \in \Lambda^{*}$, $ q_{i} \in
\mathbb{Z}$, as illustrated in figure~\ref{shifts}. Such lattice shifts
give rise to polygons wrapping the torus, and hence to further contributions
to the $N$-point function. Note that, if the $i^{th}$ D-brane has D-branes which run parallel to it,
we must modify our shift vector to $v_{i}=q_{i}gcd(\{|I_{k,i+1}| | k \in
P(i)\})\vec{L}_{i,i+1}$, where $P(i)$ is the set of all D-branes parallel to
the $i^{th}$ D-brane and $gcd$ stands for the greatest common divisor.

The global monodromy conditions are now generalised to,
\begin{equation}
\Delta _{C_{i}}X_{cl}=e^{-\pi i(\vartheta_{i}-\vartheta_{i+1})}4\sin (\pi \vartheta_{i})\sin (\pi\vartheta_{i+1})(f_{i+1}-f_{i}+v_{i})=\oint _{C_{i}}dz\partial X_{cl}(z)+\oint_{C_{i}}d\overline{z}\bar{\partial}X_{cl}(\bar{z}).
\end{equation}
Using the contour integrals~(\ref{nints}) we obtain,
\begin{equation}
\label{monosimple2}
\begin{array}{ll}
c_{i}F^{i}_{l}=f_{l+1}-f_{l}+v_{l}, & l=1,..,N-2.
\end{array}
\end{equation}
Hence, we now have,
\begin{equation}
c_{i}(x_{2},..,x_{N-2},q_{1},..,q_{N-2})=\sum_{l=1}^{N-2}(f_{l+1}-f_{l}+v(l))(F^{-1})_{i}^{l}.
\end{equation}
Our normalisation constants are now dependent on the $q_{l}$ as well as the
$x_{i}$. The $q_{l}$ define the wrapped polygons, and take values
determined by the intersection numbers of the D-branes.

\begin{figure}
\centering
\epsfig{file=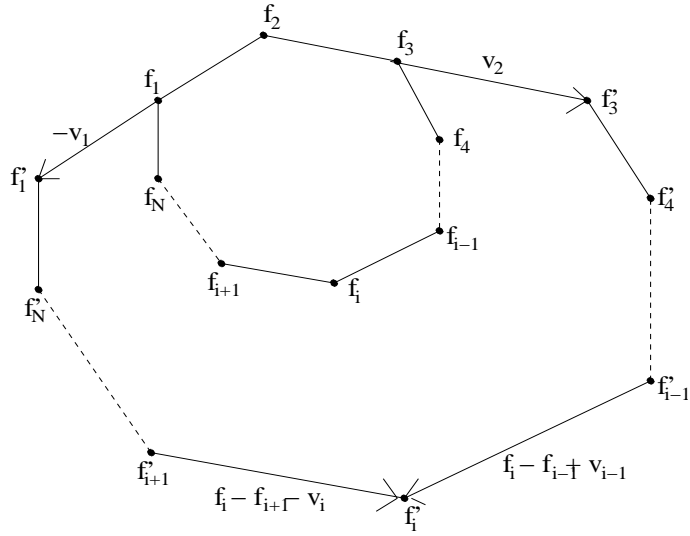, height=70mm, width=90mm}
\caption{Determining the wrapped polygons}
\label{polygonpic}
\end{figure}

To determine the wrapped polygons which contribute to the $N$-point amplitude, consider extending (w.l.o.g) from vertex
$f_{2}$ as depicted in figure~\ref{polygonpic}. Closure of the polygon requires,
\begin{equation}
\sum_{i=1}^{N}v_{i}=0.
\label{closure}
\end{equation}
Assuming there are no D-branes parallel to one another and substituting,
\begin{equation}
\label{vecs}
|I_{i,i+1}|\vec{L}_{i,i+1}=-(n_{i}v_{x}+m_{i}v_{y}),
\end{equation}
into~(\ref{closure}) we obtain the linear diophantine equations,
\begin{equation}
\left( \begin{array}{l}
q_{N}I_{N,N-1} \\
q_{N-1}I_{N,N-1}
\end{array} \right) = \left( \begin{array}{lll}
                        I_{N-1,1} & \ldots & I_{N-1,N-2} \\                        
                        I_{1,N} & \ldots & I_{N-2,N}
                           \end{array} \right) \left( \begin{array}{l}
                                                         q_{1} \\
                                                         \vdots \\
                                                         q_{N-2}
                                                       \end{array} \right).
\label{diop}
\end{equation} 
We can now determine our wrapped polygons by extending from $f_{2}$, taking all
values of $q_{i}$, $i=1,..,N-2$ that allow for integer solutions
to~(\ref{diop}). Each such solution defines a wrapped polygon contributing to
the N-point function. For example, consider the $N=3$ case. Then~(\ref{diop}) reduces to,
\begin{equation}
\begin{array}{ll}
q_{1}I_{21}=q_{3}I_{32}, & q_{2}I_{21}=q_{3}I_{13}.
\end{array}
\end{equation} 
The $q_{1}$ and $q_{2}$ which allow for integer solutions are then of the form,
\begin{equation}
\begin{array}{ll}
  q_{1}=\frac{l |I_{32}|}{gcd(|I_{21}|,|I_{32}|,|I_{13}|)}, & q_{2}=\frac{l |I_{13}|}{gcd(|I_{21}|,|I_{32}|,|I_{13}|)},
\end{array}
\end{equation}
where $l \in \mathbb{Z}$. This reproduces the result found in~\cite{paper2,mirjam} for the wrapping contributions to Yukawa interactions.

Unfortunately, it is not possible to solve~(\ref{diop}) in the general case,
only for concrete examples. This is due to the fact that diophantine
equations are usually solved algorithmically, however, it is possible to make
some general remarks. We require the following result from linear
algebra~\cite{sysdiop,algebra}, \vspace{0.25in} \newline 
THEOREM. Let $A \in M_{m,n}(\mathbb{Z})$. There exists $L \in SL_{m}(\mathbb{Z})$ and $R \in
SL_{n}(\mathbb{Z})$ such that,
\begin{equation}
LAR=D=diag(d_{1},d_{2},..,d_{s},0,..,0),
\end{equation}
where $d_{i}>0$, $i=1,..,s$ and $d_{i}|d_{i+1}$, $i=1,..,s-1$.\vspace{0.25in} \newline
Denoting our diophantine matrix equation as $b=Ax$, we can therefore write our solution as $x=Ry$ where $y$ is a solution to the diagonal
system of equations, $Dy=e$, with $e=Lb$. Since A is of size $2 \times
(N-2)$, we have in general two distinct cases for the form of D, corresponding
to $s=1$ and $s=2$ above. However, our assumption of a convex polygon means
that we only need to consider the case $s=2$. This can be deduced from the fact that
the dimension of the row space of A is the same as that of D. Hence, we have the general solution,
\begin{equation}
\left( \begin{array}{l} 
           q_{1} \\
           \vdots \\
           q_{N-2}
        \end{array} \right)=R \left( \begin{array}{l}
                                      e_{1}/d_{1} \\
                                      e_{2}/d_{2} \\
                                         l_{1} \\
                                        \vdots \\
                                         l_{N-4}
                                     \end{array} \right),
\end{equation}
where $l_{i}$ are free integer parameters and $e_{1}$ and $e_{2}$ are
functions of $q_{N}$ and $q_{N-1}$. Unfortunately, it is not possible
to determine $L$ and $R$ for a general matrix $A$. 

We can now write the classical contribution to the N-point function in the
toroidal case as,
\begin{equation}
\label{classi}
Z_{cl}(x_{2},..,x_{N-2})=\sum_{q_{1},.,q_{N-2}}e^{-Scl(x_{2},..,x_{N-2},q_{1},..,q_{N-2})},
\end{equation}
where,
\begin{equation}
\label{classicalaction}
S_{cl}(x_{2},..,x_{N-2},q_{1},..,q_{N-2})=\frac{1}{4\pi \alpha'}\left(
  |a|^{2}I(x_{2},..,x_{N-2})+\sum_{i,j}b_{i}^{*}b_{j}I'_{\bar{i}j}(x_{2},..,x_{N-2}) \right).
\end{equation}
Note that we sum over $N-2$ variables corresponding to the $N-2$ independent
sides of an $N$-sided polygon and only include $q_{i}$ such that~(\ref{diop})
has a solution. This sum incorporates all possible wrapped
polygon contributions to our classical amplitude.

\subsection{Expressing $S_{cl}$ in terms of $F_{D}$}
It is possible to express $S_{cl}$ solely in terms of generalised (type D Lauricella) hypergeometric
functions. This allows us to express $N$-point amplitudes in a more
computationally friendly manner. Firstly, since $S_{cl}$ is independent of
$x_{\infty}$, we can drop the $(y-x_{\infty})$ factors from
$I, I'_{\bar{i}j}$, and $F^{i}_{l}$ i.e. they factor out from the
integrals and cancel. Next consider, 
\begin{equation}
I(x_{2},..,x_{N-2})=\int \prod_{j=2}^{N-2}|z-x_{j}|^{-2(1-\vartheta_{i})}|z|^{-2(1-\vartheta_{1})}|z-1|^{-2(1-\vartheta_{N-1})}d^{2}z.
\end{equation}
Using the method of~\cite{kawai} to relate open and closed string amplitudes,
we can split up the above integral into a product of holomorphic and
antiholomorphic contour integrals. Thus,
\begin{equation}
\label{sclint}
I(x_{2},..,x_{N-2})=\sum_{p=2}^{N-1}(-1)^{p}\alpha_{2,p}|F^{1}_{p}||F^{1}_{1}|
+\sum_{j=2}^{N-2}\sum_{p=0}^{j-1}(-1)^{j-(p+1)}\alpha_{p+1,j}|F^{1}_{j}||F^{1}_{p}|,
\end{equation}
where $\alpha_{ij}=\sin (\pi \sum_{l=i}^{j} \vartheta_{l})$ and,
\begin{alignat}{2}
F^{1}_{N-1}&=\int_{1}^{\infty} \prod_{j=1}^{N-1} (y-x_{j})^{-(1-\vartheta_{j})}\quad  &
F^{1}_{0}&=\int_{-\infty}^{0} \prod_{j=1}^{N-1} (y-x_{j})^{-(1-\vartheta_{j})}. 
\end{alignat}

We can now relate each $F^{1}_{i}$ to a type D Lauricella hypergeometric
function. In particular, it can be easily seen that,
\begin{alignat}{1}
\label{hyperform}
F^{1}_{i} & =e^{i\pi\vartheta_{i}}(x_{i}-x_{i+1})^{-1+\vartheta_{i}+\vartheta_{i+1}}\prod_{\stackrel{j=1}{(j
  \neq
  i,i+1)}}^{N-1}(x_{i}-x_{j})^{-(1-\vartheta_{j})}B(\vartheta_{i},\vartheta_{i+1})
  \nonumber  \\
 \times &
  F_{D}^{(N-3)}(\vartheta_{i},1-\vartheta_{1},..,1-\vartheta_{i-1},1-\vartheta_{i+2},..,1-\vartheta_{N-1};\vartheta_{i}+\vartheta_{i+1};\tilde{x}_{i,1},..,\tilde{x}_{i,i-1},\tilde{x}_{i,i+2},..,\tilde{x}_{i,N-1}), 
  \nonumber \\
F^{1}_{0} & = e^{-i \pi(\vartheta_{N}+1)}B(\vartheta_{N},\vartheta_{1})\times
  \nonumber \\
\times &
  F_{D}^{(N-3)}(\vartheta_{N},1-\vartheta_{2},..,1-\vartheta_{N-2};\vartheta_{1}+\vartheta_{N};1-x_{2},..,1-x_{N-2}),
  \nonumber \\
F^{1}_{N-1} & =B(\vartheta_{N},\vartheta_{N-1})F_{D}^{(N-3)}(\vartheta_{N},1-\vartheta_{2},..,1-\vartheta_{N-2};\vartheta_{N}+\vartheta_{N-1};x_{2},..,x_{N-2}),
\end{alignat}
where $\tilde{x}_{ij}=\frac{x_{i}-x_{i+1}}{x_{i}-x_{j}}$ and $B(a,b)$ is the
beta function. The $I'_{\bar{i}j}$ can also be treated as above.

These expressions provide us with a simpler handle on the classical
contribution to the $N$-point amplitude. Let us first consider the simple case
of $N=4$. We will then see that for general $N$ and under certain
circumstances, we can minimise the classical action to the sum of the polygon areas in each of the torus subfactors. 

\subsection{The general four point amplitude} 
\begin{figure}
\begin{center}\includegraphics[%
  scale=0.5]{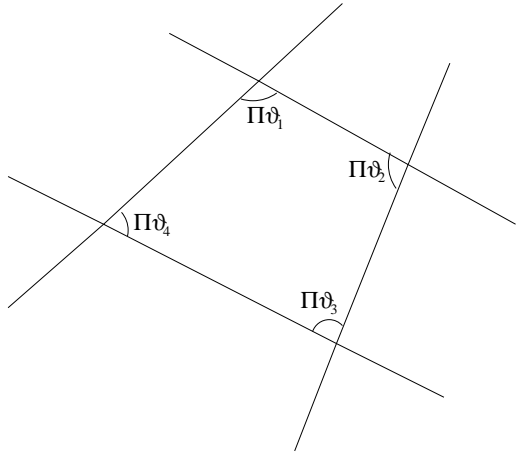}\end{center}
\caption{Generic 4 point diagram\label{cap:Generic-4-point}}
\end{figure}

Consider a four point amplitude of the general type depicted in
figure~\ref{cap:Generic-4-point}. The above expressions will enable us to
express~(\ref{4ptcls}) and the extension to wrapped polygons more
explicitly. In this case, $S_{cl}$ is simply a function of $x_{2}$ and the
wrapping numbers $q_{1}$ and $q_{2}$, and our expressions for the $F^{1}_{i}$
involve simple hypergeometric functions of the form
${}_{2}F_{1}(a,b,c;x)$. From~(\ref{hyperform}) we explicitly obtain,
\begin{equation}
\begin{array}{l}
F^{1}_{0}=e^{-i
  \pi(1+\vartheta_{4})}B(\vartheta_{4},\vartheta_{1}){}_{2}F_{1}(\vartheta_{4},1-\vartheta_{2},\vartheta_{1}+\vartheta_{4};1-x_{2}),
  \\
F^{1}_{1}=e^{-i \pi(\vartheta_{2}+\vartheta_{3})}x_{2}^{-1+\vartheta_{1}+\vartheta_{2}}B(\vartheta_{1},\vartheta_{2}){}_{2}F_{1}(\vartheta_{1},1-\vartheta_{3},\vartheta_{1}+\vartheta_{2};x_{2}),
  \\
F^{1}_{2}=e^{-i \pi(-1+\vartheta_{3})}(1-x_{2})^{-1+\vartheta_{2}+\vartheta_{3}}B(\vartheta_{2},\vartheta_{3}){}_{2}F_{1}(\vartheta_{3},1-\vartheta_{1},\vartheta_{2}+\vartheta_{3};1-x_{2}),
  \\
F^{1}_{3}=B(\vartheta_{4},\vartheta_{3}){}_{2}F_{1}(\vartheta_{4},1-\vartheta_{2},\vartheta_{4}+\vartheta_{3};x_{2}).
\end{array}
\end{equation}
Note that $F^{1}_{i}=F_{i}$ and $F^{2}_{i}=\bar{F}'_{i}$ as defined in subsection~\ref{solnsandmono}. We
can relate $F^{1}_{0}$ and $F^{1}_{3}$ to $F_{1}$ and $F_{2}$ as follows,
\begin{equation}
\begin{array}{l}
|F^{1}_{0}|=\frac{\sin(\pi \vartheta_{2})}{\sin(\pi \vartheta_{1})(1-\alpha
 \beta)}\left( |F_{2}|+\beta |F_{1}| \right), \\
|F^{1}_{3}|=\frac{\sin(\pi \vartheta_{2})}{\sin(\pi \vartheta_{3})(1-\alpha
 \beta)}\left( |F_{1}|+\alpha |F_{2}| \right),
\end{array}
\end{equation}
where, 
\begin{equation}
\begin{array}{ll}
\alpha=-\frac{\sin(\pi(\vartheta_{1}+\vartheta_{2}))}{\sin(\pi\vartheta_{1})},
&
\beta=-\frac{\sin(\pi(\vartheta_{2}+\vartheta_{3}))}{\sin(\pi\vartheta_{3})}.
\end{array}
\end{equation}
Substituting into~(\ref{sclint}) we obtain,
\begin{equation}
\label{simpleI}
I(x_{2})=\frac{\sin(\pi\vartheta_{2})}{1-\alpha \beta}(\beta|F_{1}|^{2}+2|F_{1}||F_{2}|+\alpha|F_{2}|^{2}).
\end{equation}
Furthermore, for this simple case, we can derive the relations,
\begin{equation}
\label{fprime}
\begin{array}{l}
|F'_{1}|=(1-x_{2})^{1-\vartheta_{2}-\vartheta_{3}}x_{2}^{1-\vartheta_{1}-\vartheta_{2}}\gamma(|F_{1}|+\alpha|F_{2}|), \\
 |F'_{2}|=(1-x_{2})^{1-\vartheta_{2}-\vartheta_{3}}x_{2}^{1-\vartheta_{1}-\vartheta_{2}}\gamma(|F_{2}|+\beta|F_{1}|),
\end{array}
\end{equation}
where, 
\begin{equation}
\gamma=\frac{\Gamma(1-\vartheta_{2})\Gamma(1-\vartheta_{4})}{\Gamma(\vartheta_{1})\Gamma(\vartheta_{3})}.
\end{equation}
Note that $I'$ and $F'_{i}$ are obtained from $I$ and $F_{i}$ by the
substitution $\vartheta_{i} \rightarrow 1-\vartheta_{i}$. Hence, $I'(x_{2})$ can now also be obtained in terms of $|F_{1}|$ and $|F_{2}|$ simply by letting
$\vartheta_{i} \rightarrow 1-\vartheta_{i}$ in~(\ref{simpleI}) and substituting
in~(\ref{fprime}). We also require the expressions,
\begin{equation}
\begin{array}{ll}
|a|= \left(\frac{v_{21}|F'_{2}|+v_{32}|F'_{1}|}{|F_{1}||F'_{2}|+|F_{2}||F'_{1}|} \right),
 &  |b|= \left(\frac{v_{32}|F_{1}|-v_{21}|F_{2}|}{|F_{1}||F'_{2}|+|F_{2}||F'_{1}|} \right),
\end{array}
\end{equation}
where $v_{21}=|f_{2}-f_{1}+v_{1}|$ and $v_{32}=|f_{3}-f_{2}+v_{2}|$.
This allows us to obtain the following contribution to the classical action from a single $T_{2}$,
\begin{equation}
\label{4ptaction}
S^{T_{2}}_{cl}(\tau,v_{21},v_{32})=\frac{\sin(\pi\vartheta_{2})}{4\pi\alpha'}\left(\frac{((v_{21}\tau-v_{32})^{2}+\gamma\gamma'(v_{21}(\beta+\tau)+v_{32}(1+\alpha\tau))^{2})}{(\beta+2\tau+\alpha\tau^{2})}\right),
\end{equation}
where $\tau(x_{2})=\left|\frac{F_{2}}{F_{1}}\right|$ and $\gamma'$ is
obtained from $\gamma$ by the substitution $\vartheta_{i} \rightarrow
1-\vartheta_{i}$. 

The complete expression for the action is just a sum of these
contributions, one from each torus subfactor, i.e.
\begin{equation}
\label{sumscl}
S_{cl}=\sum_{i=1}^{3}S^{T^{i}_{2}}_{cl}(\tau^{i},v^{i}_{21},v^{i}_{32}).
\end{equation}  
Now, if there is only non-zero worldsheet area in one subtorus, or if the 
angles (and hence $\tau^i$) and ratios of lengths are the same in every
subtorus (i.e. the polygons are identical up to a scaling), we may use a
saddle point approximation to minimise the complete $S_{cl}$. The minimum 
of $S^{T_{2}}_{cl}$ is given by,
\begin{equation}
\tau_{min}=\frac{v_{32}}{v_{21}},
\end{equation}
and after some manipulation we find,
\begin{equation}
S^{T_{2}}_{cl}(\tau_{min})=\frac{1}{2\pi\alpha'}\left(\frac{\sin\pi\vartheta_{1}\sin\pi\vartheta_{4}}{\sin(\pi\vartheta_{1}+\pi\vartheta_{4})}\frac{v_{14}^{2}}{2}-\frac{\sin\pi\vartheta_{2}\sin\pi\vartheta_{3}}{\sin(\pi\vartheta_{2}+\pi\vartheta_{3})}\frac{v_{23}^{2}}{2}\right).
\end{equation}
We recognize this as the area/$2\pi\alpha'$ of the four-sided
polygon. Note also that at this minimum $b=0$. Hence, in this case, we see that $S_{cl}$ is minimised to the sum of the areas of the
quadrilaterals from each $T^{2}$ subfactor. 
 
As we have seen, at the minimum of $S^{T_{2}}_{cl}$ for $N=4$, $b=0$ and we
obtain the area of the polygon involved in the interaction. An analogous
situation also occurs in the three point case, where again $b=0$ and the
action is the area of a triangle. Furthermore, it seems intuitively obvious
that for general $N$, the area of the worldsheet (i.e. $S_{cl}$) has as its
minimum value the area of the polygon associated with the amplitude, provided
only one $S^{T^{i}_{2}}_{cl}$ is non-zero. That this is indeed the case, 
and this occurs when the $b_{i}=0$, can be motivated as follows.

We can express $X_{cl}$ as,
\begin{equation}
\label{sc}
X_{cl}(z,\bar{z})=A+a\int^{z}\prod_{i=1}^{N}((\zeta
-x_{i})^{-(1-\vartheta_{i})}d\zeta+b^{*}_{l}\int^{\bar{z}} \prod_{i=1}^{N}\overline{(\bar{\zeta}
-x_{i})^{-\vartheta_{i}}}\rho^{l}(\bar{\zeta})d\bar{\zeta},
\end{equation}
with $X_{cl}(x_{i})=f_{i}$ and $A\in \mathbb{C}$. The lower integration limits are left
unspecified as they affect only the value of A. Differentiating gives
$\partial X_{cl}$ and $\bar{\partial}X_{cl}$, as in section~\ref{Nptclassical}. 
Note that we expect $X_{cl}$ to be, at least locally, one to one and hence if $f_{j}=f_{i}$ we must have $x_{j}=x_{i}$.
Using~(\ref{sc}) and $X_{cl}(x_{i})=f_{i}$ allows us immediately to obtain the relation,
\begin{equation}
\label{vertexreln}
\begin{array}{ll}
c_{l}F^{l}_{i}=f_{i+1}-f_{i} & i=1,..,N-2,
\end{array}
\end{equation}
which are simply the global monodromy conditions~(\ref{genmono}).

Now inserting $b_{i}=0$ in~(\ref{sc}), we obtain a Schwarz-Christoffel map.
This is the general form of a map from the upper-half complex plane to an $N$-sided polygon.
Hence, integrating this over the complex plane as
in~(\ref{classicalaction}) just gives back the area of the polygon as the 
classical action. In summary, we have motivated the simple rule, 
\begin{quote}
\textit{The minimum of the classical action equals the sum of the polygon
  areas projected in each $T^{2}$ when the non-zero polygons are the same up to an overall scaling.}
\end{quote}
From now on, in keeping with the literature on Schwarz-Christoffel mappings, we will refer
to the $x_{i}$ as prevertices.

\subsection{Wrapping contributions}

To completely determine the classical contribution to the $4$-point
amplitude, we must also include contributions from polygons which
wrap the tori. We need to determine what values of $q_{1}$ and $q_{2}$ to sum over to
include all the wrapping contributions to the amplitude. The integers $q_{i}$ we
require are simply those which allow for an integer solution to the
following system of two diophantine equations obtained from~(\ref{diop}),
\begin{equation}
\left( \begin{array}{l}
q_{1}I_{21} \\
q_{2}I_{21}
\end{array} \right) = \left( \begin{array}{ll}
                        I_{32} & I_{42} \\                        
                        I_{13} & I_{14}
                           \end{array} \right) \left( \begin{array}{l}
                                                         q_{3} \\
                                                         q_{4}
                                                       \end{array} \right).
\end{equation} 
For any fixed $q_{1}$ and $q_{2}$ this solution is unique, since the
determinant of the above matrix is non-zero. However, for the general case it
is not possible to solve this for our wrapped polygons, since as mentioned
earlier a solution can only be generated algorithmically. However, for the
simplest $b=0$ cases, we see that the the classical part of any
polygon contribution to an $N$-point amplitude is essentially
$e^{-\frac{1}{2\pi \alpha'}A}$, where $A$ is the area of the wrapped polygon
in the relevant $T^{2}$ torus. Hence,
the leading contribution to the $N$-point function comes from the smallest
polygon. This is the single unwrapped polygon from the planar case
corresponding to the trivial solution to~(\ref{diop}), i.e. all $q_{i}=0$.

In the situation where we have D-branes which are parallel, our diophantine
equations~(\ref{diop}) no longer apply. The necessary modifications can be
easily determined by substituting the generalised expression,
\begin{equation}
v_{i}=q_{i}gcd(\{|I_{k,i+1}| | k \in P(i)\})\vec{L}_{i,i+1},
\end{equation}
into $\sum v_{j} =0$. To illustrate this we consider the following two cases.

\subsubsection{One independent angle}
\begin{figure}
\centering
\epsfig{file=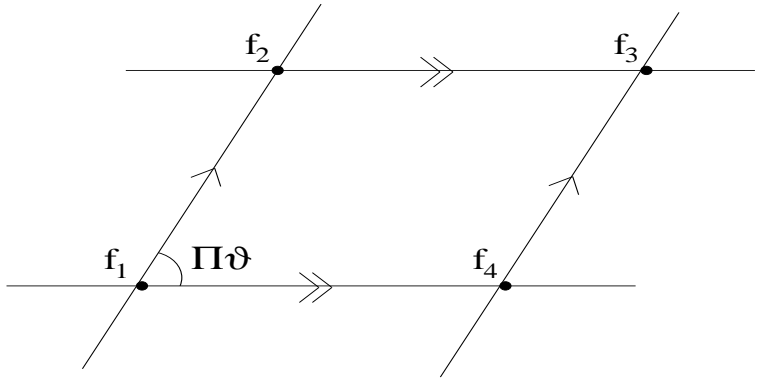, height=50mm, width=100mm}
\caption{Two sets of parallel D-branes.}
\label{oneangle}
\end{figure}

Firstly, we consider the simple case of one independent angle as depicted in figure~\ref{oneangle}.
Here we have two sets of parallel branes, the $1^{st}$ and $3^{rd}$, and the
$2^{nd}$ and $4^{th}$. As before, closure of the polygon results in the expression,
\begin{equation}
\label{4ptclos}
 \sum_{i=1}^{4} v_{i}=0.
\end{equation} 
However, we now have,
\begin{equation}
\label{vi}
v_{i}=q_{i} gcd(|I_{i,i+1}|,|I_{i+2,i+1}|)\vec{L}_{i,i+1},
\end{equation}
for $i=1,..,4$. If we define $\vec{L}_{i,i-1}$ to lie in the direction $f_{i}-
f_{i+1}$, with magnitude the distance along the $i^{th}$ brane between successive
$(i-1)^{th}$ branes, we obtain the the relation,
\begin{equation}
\label{vecrelation}
-|I_{i,i-1}|\vec{L}_{i,i-1}=|I_{i,i+1}|\vec{L}_{i,i+1}=-(n_{i}v_{x}+m_{i}v_{y}).\end{equation}
Substituting~(\ref{vi}) and~(\ref{vecrelation}) into~(\ref{4ptclos}),
we obtain the intuitively obvious result $q_{1}=q_{3}$ and
$q_{2}=q_{4}$. Hence, when summing over wrapping contributions we have a
double sum as determined previously in~\cite{paper2,mirjam}.

\subsubsection{Two independent angles} 
\begin{figure}
\centering
\epsfig{file=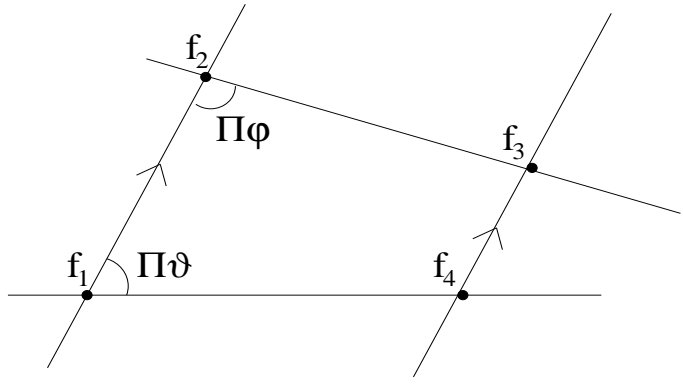, height=50mm, width=90mm}
\caption{One set of parallel D-branes.}
\label{twoangles}
\end{figure}

Next, we consider the case of two independent angles as shown in figure~\ref{twoangles}.
This time we have the $1^{st}$ and $3^{rd}$ branes parallel. Hence $v_{1}$
and $v_{3}$ are still given by~(\ref{vi}), however now $v_{2}$ and $v_{3}$
are given by the `non-parallel' expression,
$v_{i}=q_{i}|I_{i,i+1}|\vec{L}_{i,i+1}$. Again, substituting
into~(\ref{4ptclos}) and simplifying using~(\ref{vecrelation}) we obtain,
\begin{equation}
\begin{array}{ll}
q_{1}A-q_{3}A=q_{4}I_{42}, & q_{2}I_{12}=q_{4}I_{41},
\end{array}
\end{equation}
where we have defined $gcd(|I_{12}|,|I_{23}|)=a_{1}|I_{12}|+b_{1}|I_{23}|$
and $A=a_{1}I_{21}+b_{1}I_{32}$. The second expression requires,
\begin{equation}
\begin{array}{ll}
q_{2}=\frac{l I_{41}}{gcd(I_{12},I_{41})}, & q_{4}=\frac{l I_{12}}{gcd(I_{12},I_{41})},
\end{array}
\end{equation}
where $l \in \mathbb{Z}$. Defining,
\begin{equation}
c(l):=\frac{l I_{12}I_{42}}{gcd(I_{12},I_{41})}, 
\end{equation}
we obtain an infinite set of diophantine equations,
\begin{equation}
\label{infdiop}
(q_{1}-q_{3})A=c(l), 
\end{equation}
labelled by $l$. For a fixed l, a solution to~(\ref{infdiop}) exits if and
only if $A$ divides $c(l)$. In which case, we have an infinite number of solutions
given by,
\begin{equation}
\begin{array}{ll}
q_{1}=\frac{k l I_{12} I_{42}}{A gcd (I_{12},I_{41})}, & q_{3}=\frac{(-1+k)l
  I_{12}I_{42}}{A gcd(I_{12},I_{41})},
\end{array}
\end{equation}
where $k \in \mathbb{Z}$. Hence, our wrapped polygons are determined by
extending from $f_{2}$ using,
\begin{equation}
\begin{array}{ll}
q_{1}=\frac{k l I_{12} I_{42}}{A gcd (I_{12},I_{41})}, & q_{2}=\frac{l I_{41}}{gcd(I_{12},I_{41})},
\end{array}
\end{equation}
and summing over $l, k \in \mathbb{Z}$ such that $q_{1} \in \mathbb{Z}$.

\section{The quantum contribution}

\subsection{General 4-point quantum contribution}

We now turn to the quantum contribution to the amplitude, beginning
in this subsection with the generalization of the four fermion scattering
amplitude to cases where four independent branes intersect at arbitrary
angles. We wish to calculate the instanton contribution shown in figure
\ref{cap:Generic-4-point}. 

This amplitude is given by a disc diagram with four fermionic vertex
operators in the $-1/2$ picture, $V^{(a)}$ on the boundary. The
diagram can then be mapped to the upper half plane with vertices on
the real axis by using the same Schwarz-Christoffel mapping as discusses
earlier for the classical contribution. The positions of the vertices
($x_{1}\ldots x_{4}$) will eventually be fixed by $SL(2,R)$ invariance
to $0,x,1,\infty$ (where $x$ is real), so that the 4 point ordered
amplitude can be written 
\begin{equation}
\begin{array}{ll}
S_{4}(1,2,3,4) & =(2\pi)^{4}\delta^{4}(\sum_{a}k_{a})\, A(1,2,3,4) \\
& =\frac{-i}{g_{s}l_{s}^{4}}\int_{0}^{1}dx\langle
V^{(1)}(0,k_{1})V^{(2)}(x,k_{2})V^{(3)}(1,k_{3})V^{(4)}(\infty,k_{4})\rangle.
\end{array}
\end{equation}
 To get the total amplitude we sum over all possible orderings; \begin{eqnarray}
A_{total}(1,2,3,4) & = & A(1,2,3,4)+A(1,3,2,4)+A(1,2,4,3)\nonumber \\
 & + & A(4,3,2,1)+A(4,2,3,1)+A(4,3,1,2).\end{eqnarray}
 The vertex operators for the fermions are of the form \begin{equation}
V_{i}^{(a)}(x_{a},k_{a})=const\,\,\lambda^{a}u_{\alpha}^{(i)}S_{i}^{\alpha}\sigma_{\vartheta_{i}}e^{-\phi/2}e^{ik_{a}.X}(x_{a}).\end{equation}
 Here $u_{\alpha}$ is the space time spinor polarization, and $S^{\alpha}$
is the so called spin-twist operator of the form \begin{equation}
S_{i}^{\alpha}=\prod_{l=1}^{5}:\exp(iq_{i}^{l}H_{l}):\end{equation}
 where for D6 branes intersecting at angles we have $q_{i}^{l}=\left(\pm\frac{1}{2},\pm\frac{1}{2},\vartheta_{i}^{1}-\frac{1}{2},\vartheta_{i}^{2}-\frac{1}{2},\vartheta_{i}^{3}-\frac{1}{2}\right)$,
with the relative sign of the first two entries being determined by
the helicity, and $\vartheta_{i}^{m=1,2,3}$ being the angles of the
$i'$th intersection in the $m'$th complex plane. In what follows
we will frequently drop the $m$ index when we consider what is happening
in a single sub-torus. For future reference we can also identify the
available scalar fields at intersection $i$ coming from the NS sector;
their masses are \begin{eqnarray}
\alpha'M_{h}^{2} & = & \frac{1}{2}(\vartheta_{i}^{1}+\vartheta_{i}^{2}-\vartheta_{i}^{3})\nonumber \\
 &  & \frac{1}{2}(\vartheta_{i}^{1}-\vartheta_{i}^{2}+\vartheta_{i}^{3})\nonumber \\
 &  & \frac{1}{2}(-\vartheta_{i}^{1}+\vartheta_{i}^{2}+\vartheta_{i}^{3})\nonumber \\
 &  & 1-\frac{1}{2}(\vartheta_{i}^{1}+\vartheta_{i}^{2}+\vartheta_{i}^{3}).\end{eqnarray}
The spin fields have conformal dimension \begin{equation}
h_{i}=\frac{q_{i}^{2}}{2}.\end{equation}
 Here $\sigma_{\vartheta_{i}}$ is the $\vartheta$ twist operator
for the $i'$th vertex, with conformal dimension \begin{equation}
h_{i}=\frac{1}{2}\vartheta_{i}.(1-\vartheta_{i}).\end{equation}
 The evaluation of the expectation value of products of vertex operators
is straightforward apart from the factor involving the bosonic twist
operators. This calculation can be done analogously to the closed
string case \cite{dixon}. In contrast to the restricted case discussed
in ref.\cite{paper2} with only one or two independent angles, we
need to significantly modify the techniques however. In particular
in the general case (with three independent angles) the contours required
are the same Pochammer contours discussed for the classical monodromy
conditions.

We now outline the derivation. Consider the contribution from a single
complex dimension in which the branes intersect with angles $\vartheta_{i}\pi$
where $\sum\vartheta_{i}=2$ if there are no intersections. (Other
topologies are possible. For example if there is a single intersection
in the middle of the world-sheet we require $\sum_{i=left}\vartheta_{i}=\sum_{i=right}\vartheta_{i}$
where {}``left'' and {}``right'' indicate the vertices on opposing
sides of the intersection.) We begin with the asymptotic behaviour
of the Green function in the vicinity of the twist operators. Taking
account of the world-sheet boundary by adding an image charge, the
Green function can then be written \begin{equation}
G(z,w;z_{i})=g(z,w;z_{i})+g(z,\overline{w};z_{i})\end{equation}
 where $g(z,w;z_{i})$ is the usual Green function for the closed
string. It has the following asymptotics \begin{eqnarray}
g(z,w;x_{i}) & \sim & \frac{1}{(z-w)^{2}}+\op{finite}\hspace{0.7cm}z\rightarrow w\nonumber \\
 & \sim & \frac{1}{(z-x_{i})^{-\vartheta_{i}}}\hspace{1.6cm}z\rightarrow x_{i}\nonumber \\
 & \sim & \frac{1}{(w-x_{i})^{-(1-\vartheta_{i})}}\hspace{1cm}w\rightarrow x_{i}\nonumber \\
\end{eqnarray}
 and as we have seen the holomorphic fields are proportional to \begin{eqnarray}
\partial X(z) & \equiv & \omega(z)=\prod_{i}(z-x_{i})^{-(1-\vartheta_{i})}\nonumber \\
\partial\overline{X}(z) & \equiv & \omega'(z)=\prod_{i}(z-x_{i})^{-\vartheta_{i}}.\end{eqnarray}
 so that this half of the Green function may now be written generically upto
an additional term usually denoted $A$; \begin{eqnarray}
g(z,w;z_{i}) & = & \left.\omega(z)\omega'(w)\left\{ \sum_{ij}a_{ij}\frac{(z-x_{i})(z-x_{j})}{(w-x_{i})(w-x_{j})}\frac{\prod_{k}(w-x_{k})}{(z-w)^{2}}+A\right\} \right.\nonumber \\
\end{eqnarray}
 where $a_{i<j}$ and $A$ are constants. This is the most general
function with the desired conformal properties that can be written
down as prescribed in ref.\cite{bpz}. Expanding in the various limits,
we find by inspection that it has the required asymptotics if the
constants satisfy \begin{eqnarray}
\sum_{i<j}a_{ij}=1\nonumber \\
\sum_{j=i+1}^{4}a_{ij}+\sum_{j=1}^{i-1}a_{ji}=1- \vartheta_{i} &  & .\label{acond}\end{eqnarray}
 Note that summing the second of these conditions over $i$ and using
the $\sum_{i}(1-\vartheta_{i})=2$ condition automatically gives the first
condition. Of course in the end any arbitrariness in the choice of
$a_{ij}$ and $A$ must disappear from the amplitude which must be
dependent on the $\vartheta_{i}'s$ only.

Continuing, we now determine the general form of $\langle\prod_{i}\sigma_{\vartheta_{i}}\rangle$
by considering \begin{eqnarray}
\frac{\langle T(z)\prod_{i}\sigma_{\vartheta_{i}}\rangle}{\text{$\langle\prod_{i}\sigma_{\vartheta_{i}}\rangle$}} & = & \lim_{w\rightarrow z}[g_{(}z,w)-\frac{1}{(z-w)^{2}}]\nonumber \\
 & = & -\frac{1}{2}\sum_{ij}\vartheta_{i}\vartheta_{j}\frac{1}{(z-x_{i})(z-x_{j})}\nonumber \\
 &  & +\frac{1}{2}\sum_{i<j}a_{ij}\left(\frac{1}{z-x_{i}}+\frac{1}{z-x_{j}}\right)^{2}+\frac{A}{\prod_{i}(z-x_{i})}.\end{eqnarray}
 and comparing it with the OPE of $T(z)$ with the twist operator
\begin{equation}
T(z)\sigma_{\vartheta_{j}}(x_{i})\sim\frac{h_{j}}{(z-x_{i})^{2}}+\frac{\partial_{x_{i}}\sigma_{_{\vartheta_{j}}}(x_{2})}{(z-x_{i})}+\ldots\end{equation}
 The leading $(z-x_{i})^{2}$ divergences yield the correct conformal
dimension of the twist operators; \[
h_{i}=\frac{1}{2}\vartheta_{i}(1-\vartheta_{i}).\]
 Equating coefficients of $(z-x_{i})^{-1}$ (in order to preserve
generality we postpone using $SL(2,R)$ invariance to fix $(x_{1},x_{2},x_{3},x_{4})=(0,x,1,x_{\infty})$
for the moment) yields a set of differential equations for $\langle\prod_{i}\sigma_{\vartheta_{i}}\rangle$
of the form \begin{equation}
\partial_{x_{k}}\ln\left[\langle\prod_{i}\sigma_{\vartheta_{i}}\rangle\right]=\partial_{x_{k}}\ln\left[\prod_{i<j}(x_{i}-x_{j})^{a_{ij}-(1-\vartheta_{i})(1-\vartheta_{j})}\right]
+\frac{A}{\prod_{i\neq k}(x_{k}-x_{i})}.\label{differ}\end{equation}
 All that remains is to determine $A$ which can be done using monodromy
conditions for $\partial_{z}X\partial_{w}\overline{X}$. We proceed
as for the classical calculation and consider the global monodromy conditions
 arising from the two independent Pochammer loops, $C_{l=1,2}$, encircling
 the prevertices $x_{l}$ and $x_{l+1}$. From the local monodromy conditions
 for $X_{qu}$ given in~(\ref{transqu}), we see that on completing these contours
the quantum part should be left invariant,
\begin{equation}
\Delta_{C_{l}}X_{qu}=0=\oint_{C_{l}} dz \partial X_{qu} + \oint_{C_{l}} d\overline{z} \overline{\partial} X_{qu}.
\end{equation}
We now use $SL(2,R)$ invariance to fix $(x_{1},x_{2},x_{3},x_{4})=(0,x,1,x_{\infty})$.
Taking the $w\rightarrow\infty$ limit, dividing by $w'(w)$ and extracting
the leading $x_{\infty}$ contributions, the monodromy condition applied to the Green functions,
\begin{equation}
\oint_{C_l}dz \, g(z,w) + 
\oint_{C_l}d\overline{z} \, h(\overline{z},w) = 0 \, ,
\end{equation}
gives \begin{equation}
B\oint_{C_{l}}\overline{w}'(\overline{z})d\overline{z}+A\oint_{C_{l}}w(z)dz=x_{\infty}\oint_{C_{l}}\sum_{i}a_{i4}(z-x_{i})w(z)dz\end{equation}
 for both independent contours $C_{l=1,2}$, where $B$ is a constant.
Defining $G_{l}=\int F_{l}(x)dx$ and $G'_{l}=\int F'_{l}(x)dx$ we
can solve for $A$; \begin{equation}
x_{\infty}A=(1-\vartheta_{4}-a_{24})x-a_{34}-(1-\vartheta_{4})\vartheta_{2}\left[G_{1}\partial_{x}\overline{G}'_{2}-G_{2}\partial_{x}\overline{G}'_{1}\right]/J.\label{1sta}\end{equation}
 where \[
J=F_{1}\overline{F_{2}'}-F_{2}\overline{F_{1}'}.\]
 An alternative solution can be found by taking the $z\rightarrow\infty$
limit of the monodromy condition and dividing by $w(z)$; \begin{equation}
x_{\infty}A=(\vartheta_{4}-a_{13})x-a_{12}-(1-\vartheta_{2})\vartheta_{4}\left[G'_{2}\partial_{x}\overline{G_{1}}-G_{1}'\partial_{x}\overline{G_{2}}\right]/\overline{J}.\label{2nda}\end{equation}
 A different way to get the same result is to take the previous limits,
but swap interior for exterior angles. Now it is well known that integrals
over the different Pochammer contours with integrands involving $w(z)$
generate solutions to a particular hypergeometric differential equation.
In the present case the required equation is that satisfied by the
$G_{l}$ and $G_{l}'$ which (using $\partial_{x}G_{l}=F_{l}$) can
be written; \begin{eqnarray}
(1-\vartheta_{4})\vartheta_{2}G_{l} & = & x(1-x)\partial F_{l}-(\vartheta_{1}+\vartheta_{2}-1+(\vartheta_{4}-\vartheta_{2})x)F_{l}\nonumber \\
(1-\vartheta_{2})\vartheta_{4}G'_{l} & = & x(1-x)\partial F'_{l}+(\vartheta_{1}+\vartheta_{2}-1+(\vartheta_{4}-\vartheta_{2})x)F'_{l}\end{eqnarray}
 Substituting into~(\ref{1sta}) and~(\ref{2nda}) and summing yields
the desired expression for $A$; \begin{equation}
\frac{2A}{x_{\infty}x(1-x)}=\partial_{x}\ln|J|-\frac{a_{23}+a_{14}}{(1-x)}-\frac{a_{34}+a_{12}}{x},\label{finala}\end{equation}
 where \[
|J|=|F_{1}\Vert F_{2}'|+|F_{2}\Vert F_{1}'|.\]
 We shall give a closed expression for this function (in terms of
hypergeometric functions) shortly.

Finally, inserting~(\ref{finala}) into~(\ref{differ}), and using
the relations in~(\ref{acond}) gives \begin{equation}
\langle\prod_{i}\sigma_{\vartheta_{i}}\rangle=|J|^{-\frac{1}{2}}x_{\infty}^{-\vartheta_{4}(1-\vartheta_{4})}x^{\frac{1}{2}(\vartheta_{1}+\vartheta_{2}-1)-\vartheta_{1}\vartheta_{2}}(1-x)^{\frac{1}{2}(\vartheta_{2}+\vartheta_{3}-1)-\vartheta_{2}\vartheta_{3}}.\end{equation}
 Note that as promised there is no arbitrariness in the choice of
$a_{ij}$. The function $|J|$ may be evaluated as;
\begin{equation}
\begin{array}{l}
|J|= \\
\left(\frac{x^{1-\vartheta_{1}-\vartheta_{2}}}{(1-x)^{1-\vartheta_{2}-\vartheta_{3}}}\frac{\Gamma(1-\vartheta_{1})\Gamma(\vartheta_{3})}{\Gamma(\vartheta_{3}+\vartheta_{4})\Gamma(\vartheta_{2}+\vartheta_{3})}\,\,_{2}F_{1}[1-\vartheta_{1},\vartheta_{3},\vartheta_{2}+\vartheta_{3};1-x]\,_{2}F_{1}[1-\vartheta_{1},\vartheta_{3},\vartheta_{3}+\vartheta_{4};x]\right.\\
 +\,\,\,\left.\frac{(1-x)^{1-\vartheta_{2}-\vartheta_{3}}}{x^{1-\vartheta_{1}-\vartheta_{2}}}\frac{\Gamma(\vartheta_{1})\Gamma(1-\vartheta_{3})}{\Gamma(\vartheta_{1}+\vartheta_{2})\Gamma(\vartheta_{1}+\vartheta_{4})}\,\,_{2}F_{1}[\vartheta_{1},1-\vartheta_{3},\vartheta_{1}+\vartheta_{4};1-x]\,_{2}F_{1}[\vartheta_{1},1-\vartheta_{3},\vartheta_{1}+\vartheta_{2};x]\right) 
\end{array}
\end{equation}
upto an overall factor that will be absorbed into the normalization
\cite{mirjam}, and where $_{2}F_{1}$ are the standard hypergeometric
functions. This is for a sub-$T_{2}$ torus of the compactified space.
The contributions from the three complex planes should be multiplied
together with the appropriate angles $\vartheta_{i}^{m}$ for each.
As a check, note that when
there is only one independent angle, $\vartheta_{1}=\vartheta_{3}=\vartheta$
and $\vartheta_{2}=\vartheta_{4}=1-\vartheta$ the function reduces
to that found in ref.\cite{paper2}. In addition we recover the
result with two independent angles derived in ref.\cite{mirjam} by
setting $\vartheta_{1}=1-\vartheta_{2}$ and $\vartheta_{4}=1-\vartheta_{3}$.
Note that the function has crossing symmetry; it is invariant
under $\vartheta_{1}\leftrightarrow\vartheta_{3}$ and $x\leftrightarrow1-x$. Finally 
note that the entire expression is invariant if we swap interior for exterior angles, $\vartheta_i\rightarrow 
1-\vartheta_i $.

\subsection{The quantum contribution to $N$-point amplitudes}

The same procedure can be carried out for $N$-point functions following the 
techniques in ref\cite{atick} although here modified to the tree level case.
The Green function should take the form \begin{eqnarray}
g(z,w;z_{i}) & = & \left.\omega(z)\omega'(w)\left\{ \sum_{ij}a_{ij}\frac{(z-x_{i})(z-x_{j})}{(w-x_{i})(w-x_{j})}\frac{\prod_{k}(w-x_{k})}{(z-w)^{2}}+A(w)\right\} \right.\nonumber \\
\end{eqnarray}
 where $a_{i<j}$ satisfies the same condition as above, but now $A(w)$
is a function of the form \[
A(w)=\sum_{i>j>i'>j'}b_{iji'j'}\frac{\prod_{k}(w-x_{k})}{(w-x_{i})(w-x_{j})(w-x_{i'})(w-x_{j'})}\]
 where $b_{iji'j'}$ are some coefficients. (This function may look a little lobsided since it does not 
involve $z$, however this is merely a consequence of the fact that the conformal 
weights in $\gamma(z)$ already add up to 2.) We can proceed by defining a
basis for $A(z)$ as we did for dX previously; 
\begin{equation}
A(z)=\sum_{i=2}^{N-2} d_i \prod_{\stackrel{\mbox{\scriptsize $j=2$}}{\mbox{\scriptsize $(j
      \neq i$)}}}^{N-2}(z-z_{j})
\end{equation}
where $z_{i}$ is again the set of $N-3$ prevertices corresponding to the 
prevertices that are not eventually fixed by $SL(2,R)$ invariance.
It is useful in what follows to define  
\begin{equation}
g_s(z,w)=
\omega(z)\omega'(w)
\sum_{ij}a_{ij}\frac{(z-x_{i})(z-x_{j})}{(w-x_{i})(w-x_{j})}\frac{\prod_{k}(w-x_{k})}{(z-w)^{2}}\, ,
\end{equation}
and $h(\overline{z},w)$ as;
\begin{equation}
h(\overline{z},w)=\sum_{i,j=2}^{N-2}c_{ij} \overline{\Omega}'^{i}(\overline{z})\Omega'^{j}(w) \, ,
\end{equation}
where $\Omega'^{i}(z)$ is given by~(\ref{newom}). In order to write down a solution to the monodromy conditions we again require the
$(N-2)\times (N-2)$ matrix $W^{i}_{l}$ defined in~(\ref{nints}), where $l$ labels the $N-2$ independent Pochammer contours. 
(As in the classical contribution, these integrals will generate generalized hypergeometric functions which are solutions to Appell-Lauricella systems of coupled differential equations). With these definitions a solution to the global monodromy conditions with $g$ and $h$ in the prescribed form can easily be written down as follows; 
\begin{eqnarray}
g(z,w) & = & g_s(z,w) - \omega(z) \sum_l^{N-2} (W^{-1})^l_1 \oint_{C_l} dy \, g_s (y,w) \nonumber \\
h(\overline{z},w) & = & - \sum_i^{N-2} \overline{\Omega}'^{i}(\overline{z})  
\sum_l^{N-2} (W^{-1})^l_{i} \oint_{C_l} dy \, g_s (y,w).
\end{eqnarray}
We may now insert the expression for $g_s(z,w)$ and extract the singular holomorphic 
behaviour at any one of the poles, $z_{k}$, to find the holomorphic contribution;
\begin{equation}
\partial_{x_{k}}\ln
\left[
\langle\prod_{i}
\sigma_{\vartheta_{i}}\rangle\right]=
\partial_{x_{k}}\ln\left[
\prod_{i'<j'}(x_{i'}-x_{j'})^
{-(1-\vartheta_{i'})(1-\vartheta_{j'})}
\right]
\end{equation}
If we instead extract the singular antiholomorphic 
behaviour near any one of the poles, $\overline{z}_{k}$, 
(or alternatively, 
as in the four point case, evaluate the holomorphic behaviour for the diagram 
with interior and exterior angles reversed) we find
\begin{equation}
\partial_{x_{k}}\ln
\left[
\langle\prod_{i}
\sigma_{\vartheta_{i}}\rangle\right]=
\partial_{x_{k}}\ln\left[
\prod_{j\neq k}(x_{k}-x_{j})^{1-\vartheta_k}
\prod_{i'<j'}(x_{i'}-x_{j'})^
{-\vartheta_{i'}\vartheta_{j'}}
\right]
-\sum_{l=1}^{N-2}(W^{-1})^l_k \partial_{x_{k}} W^k_l.
\end{equation}
Adding the two contributions and rearranging we get the total contribution;
\begin{equation}
\begin{array}{ll}
\label{nearly}
\partial_{x_{k}}\ln
\left[
\langle\prod_{i}
\sigma_{\vartheta_{i}}\rangle\right] & =
\partial_{x_{k}}\ln\left[
\prod_{j\neq k}(x_{k}-x_{j})^{\frac{1-\vartheta_k}{2}}
\prod_{i'<j'}(x_{i'}-x_{j'})^
{\frac{1}{2}(\vartheta_{i'}+\vartheta_{j'} -1) -\vartheta_{i'}\vartheta_{j'}}
\right] \\
& - \frac{1}{2} \sum_{l=1}^{N-2}(W^{-1})^l_k \partial_{x_{k}} W^k_l   .
\end{array}
\end{equation}
All that remains is to evaluate the trailing term. In order to do this, following ref.\cite{atick}, we note that 
\begin{equation}
\partial_{x_{k}}\ln |W| = \sum_{l=1}^{N-2}(W^{-1})^l_k \partial_{x_{k}} W^k_l  + 
\sum_{j\neq k}\sum_{l=1}^{N-2}(W^{-1})^l_j \partial_{x_{k}} W^j_l  .
 \end{equation}
By comparing the singularities as $\overline{z}\rightarrow \overline{z}_{k}$ we can evaluate the second piece;
\begin{equation}
\sum_{j\neq k}\sum_{l=1}^{N-2}(W^{-1})^l_j \partial_{x_{k}} W^j_l =
\partial_{x_{k}} 
\prod_{j\neq k}(x_{k}-x_{j})^{\frac{\vartheta_k}{2}}.
\end{equation}
Finally inserting this back into eqn.\ref{nearly} 
we arrive at the desired expression for the quantum contribution;
\begin{equation}
\label{nptquantum}
\langle\prod_{i}
\sigma_{\vartheta_{i}}\rangle
= 
|W|^{-\frac{1}{2}}
\prod_{i < j}^{N-3}
(x_{i}-x_{j})^{\frac{1}{2}}
\,
\prod_{i <j}^N (x_{i}-x_{j})^
{\frac{1}{2}(\vartheta_{i}+\vartheta_{j} -1) -\vartheta_{i}\vartheta_{j}}.
\end{equation}
This is the main result of this section. One may verify that when $N=4$ it 
gives the earlier 4 point result; the $W_l^i$ matrix is now a $2\times 2$ 
matrix of the usual hypergeometric integrals in~(\ref{integrals}) 
whose determinant is proportional to $|J|$. 

\section{The complete amplitude and applications}

Together with the classical contribution, this expression may now
be used to find, for example, the full amplitude for 4 fermion interactions
on arbitrary sets of four D-branes. For example let us recap the current-current
four fermion interaction. First we collect the remaining contributions
together. These are \begin{eqnarray}
ghosts\times\langle e^{-\phi/2}(0)e^{-\phi/2}(x)e^{-\phi/2}(1)e^{-\phi/2}(x_{\infty})\rangle & = & x_{\infty}^{\frac{1}{2}}x^{-\frac{1}{4}}(1-x)^{-\frac{1}{4}}\nonumber \\
\nonumber \\\langle e^{-ip_{1}.X}e^{-ip_{2}.X}e^{-ip_{3}.X}e^{-ip_{4}.X}\rangle & = & x^{2\alpha'p_{1}.p_{2}}(1-x)^{2\alpha'p_{2}.p_{3}}\nonumber \\
\langle e^{iq_{1}.H}e^{iq_{2}.H}e^{iq_{3}.H}e^{iq_{4}.H}\rangle_{cmp} & = & \prod_{m}^{3}x_{\infty}^{\vartheta_{4}^{m}(1-\vartheta_{4}^{m})-\frac{1}{4}}x^{\vartheta_{1}^{m}\vartheta_{2}^{m}-\frac{1}{2}(\vartheta_{1}^{m}+\vartheta_{2}^{m})+\frac{1}{4}}\times\nonumber \\
 &  & (1-x)^{\vartheta_{2}^{m}\vartheta_{3}^{m}-\frac{1}{2}(\vartheta_{2}^{m}+\vartheta_{3}^{m})+\frac{1}{4}}\end{eqnarray}
 where the last piece is for the three compactified tori factors only.
The final piece comes from the uncompactified part of the fermions
and is chirality dependent. Assume that the fermions are all of the
same chirality. Then we have \begin{eqnarray}
q_{1,3} & = & (\pm\frac{1}{2},\pm\frac{1}{2},\ldots.)\nonumber \\
q_{2,4} & = & (\pm\frac{1}{2},\mp\frac{1}{2},\ldots.)\end{eqnarray}
 Calling the first two elements $\tilde{q}_{i}$ we get an additional
factor \begin{equation}
\langle e^{-iq_{1}.H}e^{-iq_{2}.H}e^{-iq_{3}.H}e^{-iq_{4}.H}\rangle_{non-cmp}=x_{\infty}^{\tilde{q}_{4}.(\tilde{q}_{1}+\tilde{q}_{2}+\tilde{q}_{3})}x^{\tilde{q}_{1}.\tilde{q}_{2}}(1-x)^{\tilde{q}_{2}.\tilde{q}_{3}}=x_{\infty}^{-\frac{1}{2}}\end{equation}
 where we have imposed $\sum_{i}^{4}q_{i}^{l}=0$ for all $l$ (note
that $\sum_{i}^{4}\vartheta_{i}^{m}=2$ ensures that this is satisfied
for the internal components), whereupon the fermions reduce to the
two spinor components of left chirality fields as follows \begin{eqnarray}
\widetilde{q_{1}} & = & -\tilde{q}_{3}=\pm(\frac{1}{2},\frac{1}{2})\nonumber \\
\widetilde{q_{2}} & = & -\tilde{q}_{4}=\pm(\frac{1}{2},-\frac{1}{2}).\end{eqnarray}
 Identifying the two $\pm$possibilities with the Weyl spinor indices
$\alpha$ of the fermions $u_{\alpha}$, we see that in $u_{\alpha}^{(1)}\overline{u}_{\dot{\beta}}^{(2)}u_{\gamma}^{(3)}\overline{u}_{\dot{\delta}}^{(4)}$
we have opposite $\alpha\gamma$ and $\dot{\beta}\dot{\delta}$ indices
which we may write as \begin{equation}
\varepsilon_{\alpha\gamma}\varepsilon_{\dot{\beta}\dot{\delta}}=-2\sigma_{\alpha\dot{\beta}}^{\mu}\overline{\sigma}_{\mu\dot{\delta}\gamma}\end{equation}
 giving the fermion factor $\overline{u}^{(1)}\gamma_{\mu}u^{(2)}\overline{u}^{(4)}\gamma^{\mu}u^{(3)}$.
Adding the other factors above we find that dependence on $\vartheta_{i}^{m}$
cancels between the bosonic twist fields and the spin-twist fields
giving \begin{equation}
\begin{array}{lll}
A(1,2,3,4) & = & -g_{s}\alpha'(\lambda^{1}\lambda^{2}\lambda^{3}\lambda^{4}+\lambda^{4}\lambda^{3}\lambda^{2}\lambda^{1})\int_{0}^{1}dx\,\, x^{-1-\alpha's}(1-x)^{-1-\alpha't}\frac{1}{\prod_{m}^{3}|J^{m}|^{1/2}}\\
 &  & \times\left[\overline{u}^{(2)}\gamma_{\mu}u^{(1)}\overline{u}^{(4)}\gamma^{\mu}u^{(3)}\right]\sum e^{-S_{cl}(x)}\end{array}\end{equation}
 where |$J^{m}|$ is the contribution from the $m'$th internal complex
dimension, we have reinstated the Chan-Paton factors, and where $s=-(k_{1}+k_{2})^{2}$,
$t=-(k_{2}+k_{3})^{2}$, $u=-(k_{1}+k_{3})^{2}$ are the usual Mandlestam
variables.

\subsection{Obtaining the $(N-1)$-point amplitude from the $N$-point amplitude}

\begin{figure}
\begin{center}
 \epsfig{file=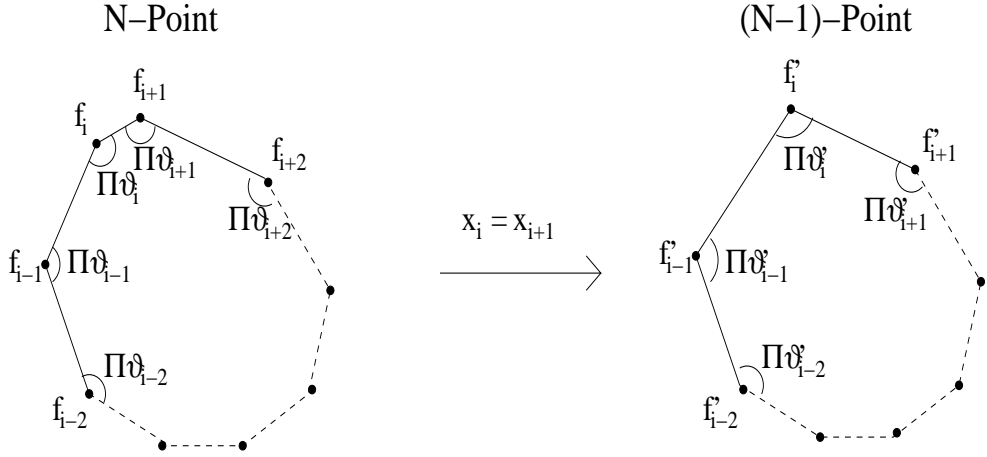,height=60mm,width=130mm}
\caption{Reducing the $N$-point amplitude to the $(N-1)$-point amplitude.}
\label{reduction}
\end{center}
\end{figure}

Consistency requires that the $(N-1)$-point amplitude be obtainable from the
$N$-point amplitude as a limiting case. Consider the set-up depicted
in figure~\ref{reduction}. We can reduce the $N$-point amplitude down to an
$(N-1)$-point amplitude as follows. Take a single prevertex, $x_{i}$ say, and let
it coalesce with $x_{i+1}$. Shifting a single prevertex of course potentially 
adjusts the entire polygon. However we may keep side 12, that is $f_{12}$, 
fixed by readjusting $a$ and keep $f_{N1}$ fixed by readjusting $b$. All $f_{j-1,j}$ 
can then be kept fixed by readjusting $x_j$ upto side $f_{i-2,i-1}$ and continuing after 
$f_{i+1,i+2}$. This operation (which is always possible) coalesces 
$f_{i}$ and $f_{i+1}$ by 
adjusting only the lengths of the three adjacent 
sides $f_{i-1,i}$, $f_{i,i+1}$, $f_{i+1,i+2}$,
whilst leaving the rest of the polygon and all the angles unchanged.

The quantum contribution in~(\ref{nptquantum}) 
should reduce in the
correct manner as two prevertices coalesce like this, and indeed it does. 
The discussion is 
similar to the one loop closed string diagram discussion of ref.\cite{atick},
so we shall just describe the salient
points. In particular when two twist fields at $x_i$ and $x_j$ come together 
we should recover an amplitude consistent with 
\begin{equation}
\label{eq:last}
\sigma_{\vartheta_i}(x_i) \sigma_{\vartheta_j}(x_j) \sim 
(x_i-x_j)^{k_{ij}} \sigma_{\vartheta_i+\vartheta_j-1}
\end{equation}
where $k_{ij}$ is to be determined by equating the conformal dimensions on each side.
We find that $k_{ij}=-(1-\vartheta_i)(1-\vartheta_j)$ when
$ \vartheta_i+\vartheta_{j} \geq 1 $ and $k_{ij}=-\vartheta_i\vartheta_j$ when 
$ \vartheta_i+\vartheta_{j} \leq 1 $. We have to verify that the amplitude yields the 
correct factors asymptotically. 
The basis of $N-2$ Pochammer loops evolves into
$N-3$ loops as two points coalesce. When the two adjacent points come together the
local behaviour of $\partial X$ is given by $\omega(z) \sim
z^{-(1-\vartheta_i)}(z-\delta)^{-(1-\vartheta_{i+1})}$ where the first 
prevertex we arbitrarily set to $0$ and the second to $\delta$. One can
check, from our expressions~(\ref{hyperform}) for the $F_{i}$,
that the $\omega $ contour integrals around the loop ``trapped'' between the two 
vertices diverge as
$(x_i-x_{i+1})^{\vartheta_i+\vartheta_{i+1}-1}$ if 
$ \vartheta_i+\vartheta_{i+1} < 1 $ 
and are convergent otherwise. The opposite is true for the $\Omega'$ integrals
so that at coalescence, including the
divergent factors from $|W|^{-1/2} \prod_{i<j}(x_{i}-x_{j})^{1/2}$, we have
\begin{equation}
\langle\prod_{j}
\sigma_{\vartheta_{j}}\rangle \sim  
(x_i-x_{i+1})^{\frac{1\pm 1}{2}(\vartheta_i+\vartheta_{i+1}-1)-
  \vartheta_i\vartheta_{i+1} }\langle\prod_{j \neq i}
\sigma_{\vartheta_{j}}\rangle
\end{equation} 
where we take a plus sign if $ \vartheta_i+\vartheta_{i+1} \geq 1 $ and minus 
if $ \vartheta_i+\vartheta_{i+1} < 1 $,
yielding the expected behaviour in~(\ref{eq:last}). 

The classical contribution to the amplitude can be dealt with in a similar fashion. 
We denote by $\bar{S}_{cl}$, the classical action for the $N$-point amplitude with
the identification $x_{i}=x_{i+1}$, and by $\tilde{S}_{cl}$ the action we would
expect for the $(N-1)$-point amplitude obtained by reduction from the $N$-point
amplitude. This notation will also be employed in this subsection to distinguish other
quantities between the two cases. We now show that
$\bar{S}_{cl}=\tilde{S}_{cl}$.

For the $(N-1)$-point case obtained by reduction from the $N$-point case
as depicted in figure~\ref{reduction}, we have
the following relations determined by the geometry of the parent amplitude,
\begin{equation}
\begin{array}{llll}
f'_{j}=f_{j} & \vartheta'_{j}=\vartheta_{j} & x'_{j}=x_{j} & \mbox{For $j=1,..,i-1$},\\
f'_{j}=f_{j+1} & \vartheta'_{j}=\vartheta_{j+1} & x'_{j}=x_{j+1} & \mbox{For $j=i+1,..,N-1$},
\end{array}
\end{equation}
and also, 
\begin{equation}
\begin{array}{lll}
f'_{i}=f_{i+1}=f_{i} & \vartheta'_{i}=-1+\vartheta_{i}+\vartheta_{i+1} & x'_{i}= x_{i+1}.
\end{array}
\end{equation}
Using these expressions it is simple to deduce that $\bar{I}=\tilde{I}$ and furthermore, 
\begin{equation}
\label{intrelns}
\begin{array}{ll}
\bar{I}'_{\bar{l}m}=\tilde{I}'_{\overline{g(l)}g(m)}, & l,m=1,..,i-1,i+1,..,N-2, \\
\bar{I}'_{\bar{i}m}=\bar{I}'_{\overline{i+1},m},
\end{array}
\end{equation}
where,
\begin{equation}
g(k)=\left\{ \begin{array}{ll}
                 k & k=1,..,i-1 \\
                k-1 & k=i+1,..,N-2.
            \end{array} \right. 
\end{equation}
We also need to examine the global monodromy conditions in the $N$-point and
$(N-1)$-point cases. For this we require the following easily obtained relations,
\begin{equation}
\begin{array}{ll}
\bar{F}^{k}_{j}=\tilde{F}^{g(k)}_{g(m)}, & k,j=1,..,i-1,i+1,..,N-2, \\
\bar{F}^{i}_{j}=\bar{F}^{i+1}_{j}, & F^{k}_{i}=0.
\end{array}
\end{equation}
Then, we can see that the monodromy conditions are the same in both cases with
the identification,
\begin{equation}
\label{conrelns}
\begin{array}{ll}
\bar{c}_{k}=\tilde{c}_{g(k)}, & k=1,..,i-1,i+2,..,N-2, \\
\bar{c}_{i+1}+\bar{c}_{i}=\tilde{c}_{i}.
\end{array}
\end{equation}
Finally, substituting~(\ref{intrelns}) and~(\ref{conrelns}) into our
expression for $\bar{S}_{cl}$ given in~(\ref{classNpt}) we obtain
$\tilde{S}_{cl}$. Hence the $N$-point amplitude reduces to that of an $(N-1)$-point amplitude as required.

\subsection{Application: Higgs exchange and flavour violation}

\begin{figure}
\begin{center}
\epsfig{file=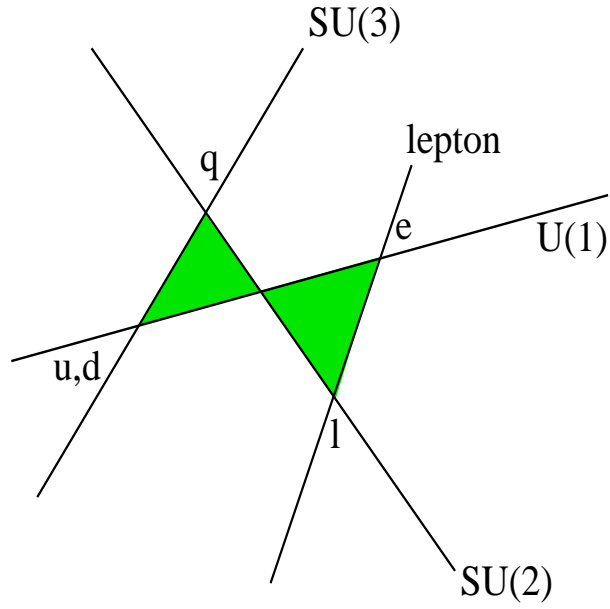,height=80mm, width=80mm}
\end{center}
\caption{t-channel Higgs exchange as {}``double instanton''}
\label{double}
\end{figure}

The general 4-point amplitude is required where there are four independent
branes. In models which reproduce the Standard Model (or supersymmetric
variants thereof) the relevant processes are therefore $q_{L}q_{R}\rightarrow e_{L}e_{R}$
as shown in figure~\ref{double}, and necessarily involve 2 left and
2 right chirality fields.

The higgs field appears at the $SU(2)$-$U(1)$ brane intersection
and so this process is the equivalent of t-channel Higgs exchange
in the field theory limit. Indeed since the instanton suppression
goes as $e^{-Area/2\pi\alpha'}$ we expect to find the product of
two Yukawas. The amplitude should go as \[
\frac{Y_{u}Y_{e}}{t-M_{h}^{2}}\]
or the $s$ channel equivalent, and we shall shortly verify the appearance
of the Higgs pole and the correctly normalized Yukawas. 

However it is also interesting to note that, merely as a result of
geometry, there can be significantly larger contributions than one
would expect in field theory. This is because the Higgs exchange involves
the product of two Yukawa couplings. In our case this would only be
the case if the lepton brane was lying to the right of the intersection
(in the figure) at which the higgs is located. Then the area of the
4 point instanton is indeed the sum of the two Yukawa triangles, and
the product is what appears in the amplitude. If on the other hand
the lepton brane is lying to the left of the intersection, the contribution
goes like $Y_{u}/Y_{e}$ and can be significantly enhanced for low
string scales. In the (unrealistic) limit that the lepton brane is
lying on top of the SU(3) brane in all sub-$T_{2}$tori, there is
no Yukawa suppression at all in this process. Note however that there
should be an overall stringy suppression as there is no field theory
limit and therefore no pole. Thus one expects a contribution like
\[
\frac{Y_{u}/Y_{e}}{M_{s}^{2}}\]
We shall now show how this behaviour emerges from the amplitude. To
anticipate the calculation, when the situation is as in figure~\ref{double},
the action is minimized in the limit $x\rightarrow1$ corresponding
to the field theory limit. In this limit we will recover the Higgs
pole with the required mass from the $x$-integral. If there is no
Higgs intersection then the action is minimized for finite $0<x<1$.
In the simplest cases we can use a saddle point approximation 
as we did earlier to obtain an expression
that goes as the area of the polygon
with a $1/M_{s}^{2}$ suppression as expected. However generically 
such a simple saddle point approximation will not exist, and the 
$\tau^m$ parameters in each sub-torus will take different values. In this 
case interesting flavour structure can emerge.

\begin{figure}
\begin{center}
\epsfig{file=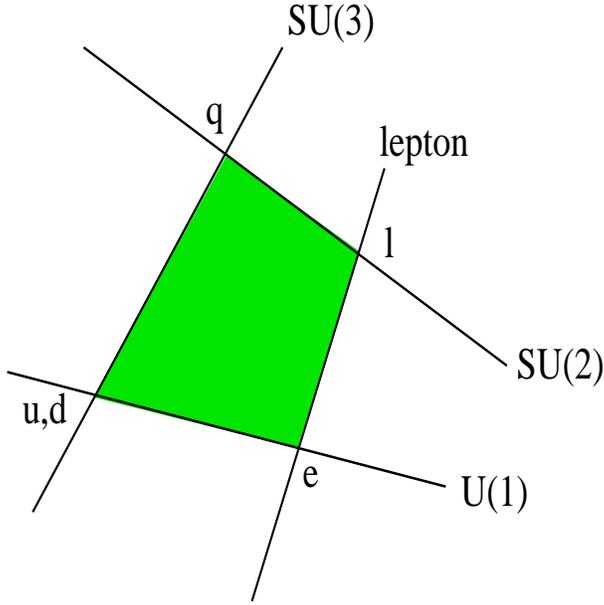,height=80mm,width=80mm}
\end{center}
\caption{Equivalent instanton contribution to flavour changing}
\label{stri}
\end{figure}
In order to show how this behaviour arises we first determine the
amplitude. We are interested in the operator $(\overline{q}_{L}^{(3)}q_{R}^{(2)})(\overline{e}_{R}^{(1)}e_{L}^{(4)})$
for which the uncompactified part of the fermions now have charges
\begin{eqnarray}
\widetilde{q_{1}} & = & -\tilde{q}_{4}=\pm(\frac{1}{2},\frac{1}{2})\nonumber \\
\widetilde{q_{2}} & = & -\tilde{q}_{3}=\pm(\frac{1}{2},-\frac{1}{2}).\end{eqnarray}
 We get slightly different factors from the current-current operator.
In particular \begin{eqnarray}
\langle e^{-iq_{1}.H}e^{-iq_{2}.H}e^{-iq_{3}.H}e^{-iq_{4}.H}\rangle_{non-cmp} & = & x_{\infty}^{\tilde{q}_{4}.(\tilde{q}_{1}+\tilde{q}_{2}+\tilde{q}_{3})}x^{\tilde{q}_{1}.\tilde{q}_{2}}(1-x)^{\tilde{q}_{2}.\tilde{q}_{3}}\nonumber \\
 & = & x_{\infty}^{-\frac{1}{2}}x^{\tilde{q}_{1}.\tilde{q}_{2}}(1-x)^{\tilde{q}_{2}.\tilde{q}_{3}}=x_{\infty}^{-\frac{1}{2}}(1-x)^{-\frac{1}{2}}.\end{eqnarray}
 The additional half-integer power in the $t$-channel looks unusual,
but it will make sense when we extract the Higgs pole in the $x\rightarrow1$
limit. An additional difference compared to the current-current case
is that particles of one chirality take the opposite Weyl index to
that of the opposing chirality. Using Weyl notation, in $\overline{u}_{\dot{\alpha}}^{(3)}\overline{u}_{\dot{\beta}}^{(2)}u_{\gamma}^{(1)}u_{\delta}^{(4)}$
we now have opposite $\dot{\alpha}\dot{\beta}$ and $\gamma\delta$
indices which (writing as $\varepsilon_{\dot{\alpha}\dot{\beta}}\varepsilon_{\gamma\delta}$)
just contracts the $\overline{q}_{L}q_{R}$ and $\overline{e}_{L}e_{R}$
fermions. The final expression for the amplitude is (in four component
notation) simply \begin{equation}
\begin{array}{lll}
A(1,2,3,4) & = & -g_{s}\alpha'\,\int_{0}^{1}dx\,\, x^{-1-\alpha's}(1-x)^{-1-\alpha't}\frac{(1-x)^{-\frac{1}{2}}}{\prod_{m}^{3}|J^{m}|^{1/2}}\\
 &  & \,\,\,\,\times\left[(\overline{u}^{(3)}u^{(2)})(\overline{u}^{(1)}u^{(4)})\right]\sum e^{-S_{cl}(x)}\end{array}\end{equation}
The classical action is given by~(\ref{sumscl}) in terms of the different 
$\tau$ in each sub-torus. 
It is easy to show that when the diagram has an intersection in a particular 
sub-torus, as in
figure~\ref{double}, the contribution to the action from that subtorus 
is a monotonically decreasing 
function of $x$. Hence the sum of the contributions is minimized 
by taking the limit $x\rightarrow1$.
Assuming that $1-\vartheta_{2}-\vartheta_{3}>0$, the relevant limits
are\begin{eqnarray}
Lim_{x\rightarrow1}(\tau) & = & -\beta\nonumber \\
Lim_{x\rightarrow1}(J) & = & (1-x)^{(-1+\vartheta_{2}+\vartheta_{3})}\,\frac{1}{\gamma}\,\,\eta(\vartheta_{2},\vartheta_{3})\eta(1-\vartheta_{1},1-\vartheta_{4})\end{eqnarray}
where \begin{equation}
\eta(\vartheta_{i},\vartheta_{j})=\left(\frac{\Gamma(\vartheta_{i})\Gamma(\vartheta_{j})\Gamma(1-\vartheta_{i}-\vartheta_{j})}{\Gamma(1-\vartheta_{i})\Gamma(1-\vartheta_{j})\Gamma(\vartheta_{i}+\vartheta_{j})}\right)^{\frac{1}{2}}\end{equation}
 The normalization of the amplitudes and Yukawas can be obtained in
this limit as in ref.\cite{mirjam}. We take the limit where the 4-point function with no intersection $\rightarrow$ the $3$-point. The
normalization factor for the general $4$-point function is \begin{equation}
2\pi\prod_{m=1}^{3}\sqrt{\frac{4\pi}{\gamma_{m}}\,\frac{\eta(1-\vartheta_{2}^{m},1-\vartheta_{3}^{m})}{\eta(\vartheta_{1}^{m},\vartheta_{4}^{m})}}\end{equation}
and the Yukawas take the form found in ref\cite{mirjam};\begin{equation}
Y_{23}(A_{m})=16\pi^{\frac{5}{2}}\prod_{m}^{3}\eta(1-\vartheta_{2}^{m},1-\vartheta_{3}^{m})\,\sum_{m}e^{-A_{m}/2\pi\alpha'}\end{equation}
where $A_{m}$ is the projected area of the triangles in the $m$'
th 2-torus. Once we add the intersection the interior $\vartheta_{1},\,\vartheta_{4}$
angles become exterior and should be replaced by $1-\vartheta_{1},\,1-\vartheta_{4}$
respectively. The constraint on the interior angles is now $\vartheta_{1}+\vartheta_{4}=\vartheta_{2}+\vartheta_{3}$
because of the intersection. Looking at $Lim_{x\rightarrow1}(J)$
we see that this can be taken into account by adding an extra $\sqrt{\eta(1-\vartheta_{1},1-\vartheta_{4})/\eta(\vartheta_{1},\vartheta_{4})}=\eta(1-\vartheta_{1},1-\vartheta_{4})$
factor to the $4$-point amplitude. We then find\begin{equation}
S_{4}=\alpha'\, Y_{23}(0)\, Y_{14}(0)\, e^{-S_{cl}(1)}\,\int_{0}^{1}dx\,(1-x)^{-\alpha't-\sum\frac{1}{2}(\vartheta_{2}^{m}+\vartheta_{3}^{m})}\end{equation}
The contribution to the
classical action from each sub-torus becomes\begin{equation}
S_{cl}(1)=\frac{1}{2\pi\alpha'}\left(\frac{\sin\pi\vartheta_{1}\sin\pi\vartheta_{4}}{\sin(\pi\vartheta_{2}+\pi\vartheta_{3})}\frac{v_{14}^{2}}{2}+\frac{\sin\pi\vartheta_{2}\sin\pi\vartheta_{3}}{\sin(\pi\vartheta_{2}+\pi\vartheta_{3})}\frac{v_{23}^{2}}{2}\right)\end{equation}
Bearing in mind that the angle at the intersection is $\pi-\pi\vartheta_{2}-\pi\vartheta_{3}$,
we see that the result is just the sum of the area/$2\pi\alpha'$
of the two triangles. Finally the pole term now arises from the $x$
integral;\begin{equation}
\alpha'\int_{0}^{1}dx\,(1-x)^{-\alpha't-\sum\frac{1}{2}(\vartheta_{2}^{m}+\vartheta_{3}^{m})}=\frac{1}{t-M_{h}^{2}}\end{equation}
where (recalling that $0<\vartheta_{2}^{m}+\vartheta_{3}^{m}<1$)
we recognize the mass of a scalar Higgs state in the spectrum at the
intersection;\begin{equation}
\alpha'M_{h}^{2}=1-\frac{1}{2}\sum_{m}(\vartheta_{2}^{m}+\vartheta_{3}^{m}).\end{equation}
 The opposite ordering of operators leads to the $s$- channel exchange
in the $x\rightarrow0$ limit. The above discussion was carried out
for intersecting D6-branes, but it is straightforward to translate
it to other set-ups. 

Having verified the expected field theory limits, we now turn to the
more interesting case of stringy processes that have no field theory
equivalent, where there is no Higgs intersection as in figure~\ref{stri}.
Let us begin with a remark about the typical set-up. 
One thing to notice about them is that the three generations
of left-handed quarks live at different intersections in sub $T_{2}$
tori where the right handed fields overlap, and vice versa. We will
refer to these tori as the {}``left'' and {}``right'' tori respectively.
The Yukawa couplings are therefore factorizable (since it is the areas
projected in the $T_{2}$ tori that appear in the action)\begin{equation}
(Y_{q})_{ij}=\omega_{i}\omega'_{j}\end{equation}
This is a problem for phenomenology since there are two massless eigenstates.
Clearly the factorizability remains in the above Higgs exchange process,
however generically
the stringy $4$-point couplings induce terms that do not factorize,
and this can be an important contribution to the flavour structure.

There have been attempts to overcome the factorizability problem
directly~\cite{khalil}, however it may be that the necessary flavour
structure can be introduced by four point couplings. To see this 
consider the contribution to $(\overline{q}_{L_{i}}q_{R_{j}})(\overline{q}_{R_{j'}}q_{L_{i'}})$
(where $i,j,i',j'$ label generation) amplitudes that have a non-zero
area in only one sub $T_{2}$ torus (i.e. that change flavour in either
the left or the right-handed fields but not both). In the case that
there is no intersection the action has a minimum and we can take
a saddle point approximation as we did previously and get simply 
the area/$2\pi\alpha'$ of the four-sided polygon.
This {}``trivial'' contribution is proportional to $Y_{ij}/Y_{i'j}=a_{i}/a_{i'}$
and is independent of the right-handed field. Hence such contributions
cannot change the rank of the Yukawa couplings via radiative corrections.
Factorizability is completely removed however when both left and right
fields change generation number because now there are two conflicting
contributions to the classical action and \textit{the saddle point
approximation no longer simply gives the sums of areas in the sub
tori.} The one {}``degenerate'' exception that we saw earlier 
is when the world-sheet
instanton is exactly the same in both tori (i.e. when the polygons in each
sub-tori are identical up to an overall scaling). Note however
that (since there are three generations) there must always be some
flavour changing processes which are neither trivial nor degenerate. 
This type of effect can be utilized to create a more realistic flavour 
structure and will be investigated further in ref.\cite{steve2}.

\section{Conclusion}

In summary, we have analysed the $N$-point amplitude at tree level for open strings localised
at D-brane intersections. We were able to generalise the techniques used for four point 
amplitudes on restricted configurations in~refs.\cite{paper2,mirjam}, to 
obtain both the classical
and quantum contributions to the general $N$ point amplitude, 
including those contributions that wrap the internal
space. We have also shown
that for general $N$, \textit{the minimum of the
  classical action is given by the sum of the polygon areas projected in each torus
  subfactor, when the non-zero polygons are identical up to an overall
  scaling}.

Our results are applicable to all orbifold, orientifold and toroidal
compactifications. Orbifolds and orientifolds would require a modification of
the counting over wrapped polygons, but the leading contributions would be
the same.

General $N=4$  ``contact interactions'' on four independent sets 
of D-branes were discussed, of relevance to processes such as $q_{L}q_{R}
\rightarrow e_{L}e_{R}$. This process, depending as it does on 
the geometry of the
D-branes, has a purely stringy contribution,
as well as sensible field theory limit contributions 
corresponding to $s$ and $t$-channel  Higgs exchange. 
In the former case, the amplitude is of the form
$\frac{Y_{q}/Y_{e}}{M_{s}^{2}}$ which should be 
compared to the field theory case $\frac{Y_{q}Y_{e}}{t-m_{h}^{2}}$. 
Thus at low string scales the stringy exchanges are potentially important. 
Finally, we discussed the reduction of the $N$-point amplitude down
 to the $(N-1)$-point amplitude.

These calculations provide a starting point for discussing general interactions in
intersecting brane models and hence understanding their phenomenology in more
detail. In addition they may prove useful in addressing questions to do with
the possible introduction of a realistic flavour structure in these models. 
Further processes including for example Higgstrahlung (which is in 
principle a 6 point diagram) can also be treated. 
These issues will be discussed further in ref.\cite{steve2}.

\section{Acknowledgements}
It is a pleasure to thank Mirjam Cveti\v{c}, Oleg Lebedev and Jose Santiago for
discussions. This work was funded by a PPARC studentship, and by 
Opportunity Grant PPA/T/S/1998/00833.

\bibliographystyle{unsrt}
\bibliography{CPbib}

\end{document}